\documentclass[11pt]{article}
\pdfoutput=1

\usepackage{jcappub}
\usepackage{lineno}
\usepackage{color, xcolor, appendix,latexsym,amsmath,amssymb,graphicx,booktabs,epsfig,hyperref,url,nicefrac,diagbox,relsize,longtable,comment,makecell,multirow,enumitem,siunitx,placeins,orcidlink}

\usepackage{dsfont}
\usepackage[english]{babel}
\usepackage[utf8]{inputenc}
\usepackage{letltxmacro}
\LetLtxMacro{\originaleqref}{\eqref}
\renewcommand{\eqref}{Eq.~\originaleqref}
\addto\extrasenglish{%
  \renewcommand{\figurename}{Figure}%
  \renewcommand{\tablename}{Table}%
}

\numberwithin{equation}{section}

\definecolor{MyBlue}{rgb}{0.15,0.15,0.70}
\hypersetup{
colorlinks=true,
citecolor=MyBlue,
linkcolor=MyBlue,
urlcolor=MyBlue
}
\definecolor{lightgray}{gray}{0.9}

\newcommand{\geneva}{D\'epartement de Physique Th\'eorique, Universit\'e de Gen\`eve, 24 quai Ernest Ansermet, 1211 Gen\`eve 4, Switzerland}
\newcommand{\gwsc}{Gravitational Wave Science Center (GWSC), Universit\'e de Gen\`eve, CH-1211 Geneva, Switzerland}
\newcommand{\mrs}{Aix-Marseille Universit\'e, Universit\'e de Toulon, CNRS, CPT, Marseille, France}

\title{Comparison of global networks of third-generation gravitational-wave detectors}

\author[a,b,1]{Michele Maggiore\,\orcidlink{0000-0001-7348-047X}\note{Corresponding author.}}
\author[a,b]{Francesco Iacovelli\,\orcidlink{0000-0002-4875-5862},}
\author[a,b]{Enis Belgacem\,\orcidlink{0000-0003-4920-0911},}
\author[c]{Michele Mancarella\,\orcidlink{0000-0002-0675-508X},}
\author[a,b]{Niccol\`o Muttoni\,\orcidlink{0000-0002-4214-2344}\,}

\affiliation[1]{\geneva}
\affiliation[2]{\gwsc}
\affiliation[3]{\mrs}

\emailAdd{Michele.Maggiore@unige.ch}

\abstract{We study the performances of a world-wide network made by a
European third-generation gravitational-wave (GW) detector, together with a 40-km Cosmic Explorer detector in the US, considering three scenarios for the European detector: (1)
Einstein Telescope (ET) in its 10-km triangle configuration; (2) ET in its   configuration featuring two 15-km L-shaped detectors in different sites, still taken to have all other ET characteristics (underground, and with each detector made of  a high-frequency interferometer and a cryogenic low-frequency interferometer);
(3) A single  L-shaped underground interferometer with the ET sensitivity curve, either with 15~km or with 20~km arm length.
Overall, we find that, 
if a  configuration with two widely separated L-shaped detectors (``2L") should be retained for ET,  
the  network made by a single-L European underground detector together with CE-40km  could already provide a very interesting intermediate step toward the construction of a full 2L+CE network, and is in any case superior to a  10-km triangle not inserted in an international network. We also study the performance of a network made  by a  single  L-shaped underground interferometer with the ET sensitivity curve together with a single 40km~CE and with  LIGO-India (taken at A$^\#$ sensitivity), and we find that it also has very interesting performances.
}

\begin{document}


\newcommand{\mmin}{m_{\rm min}}
\newcommand{\mmax}{m_{\rm max}}

\newcommand{\dgw}{d_L^{\,\rm gw}}
\newcommand{\dem}{d_L^{\,\rm em}}
\newcommand{\dcom}{d_{\rm com}}
\newcommand{\dmax}{d_{\rm max}}
\newcommand{\hatO}{\hat{\Omega}}
\newcommand{\ngw}{n_{\rm GW}}
\newcommand{\ngal}{n_{\rm gal}}

\newcommand{\red}{\textcolor{red}} 

\newcommand{\blue}{\textcolor{blue}}
\newcommand{\green}{\textcolor{green}}
\newcommand{\cyan}{\textcolor{cyan}}
\newcommand{\magenta}{\textcolor{magenta}}
\newcommand{\yellow}{\textcolor{yellow}}

\newcommand{\hc}{{\cal H}}

\newcommand{\nn}{\nonumber}

\newcommand{\Lrr}{\Lambda_{\rm\scriptscriptstyle RR}}
\newcommand{\scDE}{{\textsc{DE}}}
\newcommand{\scR}{{\textsc{R}}}
\newcommand{\scM}{{\textsc{M}}}

\newcommand{\hla}{\hat{\lambda}}
\newcommand{\iBox}{\Box^{-1}}
\newcommand{\Stu}{St\"uckelberg }
\newcommand{\phib}{\bar{\phi}}

\newcommand{\Fmn}{F_{\mu\nu}}
\newcommand{\FMN}{F^{\mu\nu}}
\newcommand{\Am}{A_{\mu}}
\newcommand{\An}{A_{\nu}}
\newcommand{\Amu}{A_{\mu}}
\newcommand{\Anu}{A_{\nu}}
\newcommand{\AMU}{A^{\mu}}
\newcommand{\AN}{A^{\nu}}
\newcommand{\ANU}{A^{\nu}}

\renewcommand\({\left(}
\renewcommand\){\right)}
\renewcommand\[{\left[}
\renewcommand\]{\right]}
\newcommand\del{{\mbox {\boldmath $\nabla$}}}
\newcommand\n{{\mbox {\boldmath $\nabla$}}}
\newcommand{\ra}{\rightarrow}

\def\lsim{\raise 0.4ex\hbox{$<$}\kern -0.8em\lower 0.62
ex\hbox{$\sim$}}

\def\gsim{\raise 0.4ex\hbox{$>$}\kern -0.7em\lower 0.62
ex\hbox{$\sim$}}

\def\lbar{{\hbox{$\lambda$}\kern -0.7em\raise 0.6ex
\hbox{$-$}}}

\newcommand\eq[1]{eq.~(\ref{#1})}
\newcommand\eqs[2]{eqs.~(\ref{#1}) and (\ref{#2})}
\newcommand\Eq[1]{Equation~(\ref{#1})}
\newcommand\Eqs[2]{Equations~(\ref{#1}) and (\ref{#2})}
\newcommand\Eqss[3]{Equations~(\ref{#1}), (\ref{#2}) and (\ref{#3})}
\newcommand\eqss[3]{eqs.~(\ref{#1}), (\ref{#2}) and (\ref{#3})}
\newcommand\eqsss[4]{eqs.~(\ref{#1}), (\ref{#2}), (\ref{#3})
and (\ref{#4})}
\newcommand\eqssss[5]{eqs.~(\ref{#1}), (\ref{#2}), (\ref{#3}),
(\ref{#4}) and (\ref{#5})}
\newcommand\eqst[2]{eqs.~(\ref{#1})--(\ref{#2})}
\newcommand\Eqst[2]{Eqs.~(\ref{#1})--(\ref{#2})}
\newcommand\pa{\partial}
\newcommand\p{\partial}
\newcommand\pdif[2]{\frac{\pa #1}{\pa #2}}
\newcommand\pfun[2]{\frac{\delta #1}{\delta #2}}

\newcommand\ee{\end{equation}}
\newcommand\be{\begin{equation}}
\def\bea{\begin{array}}
\def\eea{\end{array}}\def\ea{\end{array}}
\newcommand\ees{\end{eqnarray}}
\newcommand\bees{\begin{eqnarray}}
\def\nn{\nonumber}
\newcommand\sub[1]{_{\rm #1}}
\newcommand\su[1]{^{\rm #1}}

\def\v#1{\hbox{\boldmath$#1$}}
\def\vepsilon{\v{\epsilon}}
\def\vPhi{\v{\Phi}}
\def\vomega{\v{\omega}}
\def\vsigma{\v{\sigma}}
\def\vmu{\v{\mu}}
\def\vxi{\v{\xi}}
\def\vpsi{\v{\psi}}
\def\vth{\v{\theta}}
\def\vphi{\v{\phi}}
\def\vchi{\v{\chi}}

\newcommand{\om}{\omega}
\newcommand{\Om}{\Omega}

\def\f{\phi}
\def\D{\Delta}
\def\a{\alpha}
\def\b{\beta}
\def\ab{\alpha\beta}

\def\s{\sigma}
\def\g{\gamma}
\def\G{\Gamma}
\def\d{\delta}
\def\Si{\Sigma}
\def\eps{\epsilon}
\def\veps{\varepsilon}
\def\Ups{\Upsilon}
\def\Upsun{{\Upsilon}_{\odot}}

\def\dslash{\hspace{-1mm}\not{\hbox{\kern-2pt $\partial$}}}
\def\Dslash{\not{\hbox{\kern-2pt $D$}}}
\def\pslash{\not{\hbox{\kern-2.1pt $p$}}}
\def\kslash{\not{\hbox{\kern-2.3pt $k$}}}
\def\qslash{\not{\hbox{\kern-2.3pt $q$}}}


\newcommand{\vac}{|0\rangle}
\newcommand{\cav}{\langle 0|}
\newcommand{\hint}{H_{\rm int}}
\newcommand{\va}{{\bf a}}
\newcommand{\vb}{{\bf b}}
\newcommand{\vp}{{\bf p}}
\newcommand{\vq}{{\bf q}}
\newcommand{\vk}{{\bf k}}
\newcommand{\vx}{{\bf x}}
\newcommand{\xp}{{\bf x}_{\perp}}
\newcommand{\vy}{{\bf y}}
\newcommand{\vz}{{\bf z}}
\newcommand{\vu}{{\bf u}}

\def\p1{{\bf p}_1}
\def\p2{{\bf p}_2}
\def\k1{{\bf k}_1}
\def\k2{{\bf k}_2}

\newcommand{\emn}{\eta_{\mu\nu}}
\newcommand{\ers}{\eta_{\rho\sigma}}
\newcommand{\emr}{\eta_{\mu\rho}}
\newcommand{\ens}{\eta_{\nu\sigma}}
\newcommand{\ems}{\eta_{\mu\sigma}}
\newcommand{\enr}{\eta_{\nu\rho}}
\newcommand{\eMN}{\eta^{\mu\nu}}
\newcommand{\eRS}{\eta^{\rho\sigma}}
\newcommand{\eMR}{\eta^{\mu\rho}}
\newcommand{\eNS}{\eta^{\nu\sigma}}
\newcommand{\eMS}{\eta^{\mu\sigma}}
\newcommand{\eNR}{\eta^{\nu\rho}}
\newcommand{\ema}{\eta_{\mu\alpha}}
\newcommand{\emb}{\eta_{\mu\beta}}
\newcommand{\ena}{\eta_{\nu\alpha}}
\newcommand{\enb}{\eta_{\nu\beta}}
\newcommand{\eab}{\eta_{\alpha\beta}}
\newcommand{\eAB}{\eta^{\alpha\beta}}

\newcommand{\gmn}{g_{\mu\nu}}
\newcommand{\grs}{g_{\rho\sigma}}
\newcommand{\gmr}{g_{\mu\rho}}
\newcommand{\gns}{g_{\nu\sigma}}
\newcommand{\gms}{g_{\mu\sigma}}
\newcommand{\gnr}{g_{\nu\rho}}
\newcommand{\gsn}{g_{\sigma\nu}}
\newcommand{\gsm}{g_{\sigma\mu}}
\newcommand{\gMN}{g^{\mu\nu}}
\newcommand{\gRS}{g^{\rho\sigma}}
\newcommand{\gMR}{g^{\mu\rho}}
\newcommand{\gNS}{g^{\nu\sigma}}
\newcommand{\gMS}{g^{\mu\sigma}}
\newcommand{\gNR}{g^{\nu\rho}}
\newcommand{\gLR}{g^{\lambda\rho}}
\newcommand{\gSN}{g^{\sigma\nu}}
\newcommand{\gSM}{g^{\sigma\mu}}
\newcommand{\gAB}{g^{\alpha\beta}}
\newcommand{\gab}{g_{\alpha\beta}}

\newcommand{\gBmn}{\bar{g}_{\mu\nu}}
\newcommand{\gBrs}{\bar{g}_{\rho\sigma}}
\newcommand{\gBMN}{\bar{g}^{\mu\nu}}
\newcommand{\gBRS}{\bar{g}^{\rho\sigma}}
\newcommand{\gBMS}{\bar{g}^{\mu\sigma}}
\newcommand{\gBAB}{\bar{g}^{\alpha\beta}}
\newcommand{\gBma}{\bar{g}_{\mu\alpha}}
\newcommand{\gBnb}{\bar{g}_{\nu\beta}}
\newcommand{\gBab}{\bar{g}_{\a\b}}
\newcommand{\gbmn}{\bar{g}_{\mu\nu}}
\newcommand{\gbrs}{\bar{g}_{\rho\sigma}}
\newcommand{\gbMN}{\bar{g}^{\mu\nu}}
\newcommand{\gbRS}{\bar{g}^{\rho\sigma}}
\newcommand{\gbMS}{\bar{g}^{\mu\sigma}}
\newcommand{\gbAB}{\bar{g}^{\alpha\beta}}
\newcommand{\gbma}{\bar{g}_{\mu\alpha}}
\newcommand{\gbnb}{\bar{g}_{\nu\beta}}
\newcommand{\gbab}{\bar{g}_{\a\b}}

\newcommand{\hmn}{h_{\mu\nu}}
\newcommand{\hrs}{h_{\rho\sigma}}
\newcommand{\hmr}{h_{\mu\rho}}
\newcommand{\hns}{h_{\nu\sigma}}
\newcommand{\hms}{h_{\mu\sigma}}
\newcommand{\hnr}{h_{\nu\rho}}
\newcommand{\hrn}{h_{\rho\nu}}
\newcommand{\hra}{h_{\rho\alpha}}
\newcommand{\hsb}{h_{\sigma\beta}}
\newcommand{\hma}{h_{\mu\alpha}}
\newcommand{\hna}{h_{\nu\alpha}}
\newcommand{\hmb}{h_{\mu\beta}}
\newcommand{\has}{h_{\alpha\sigma}}
\newcommand{\hab}{h_{\alpha\beta}}
\newcommand{\hnb}{h_{\nu\beta}}
\newcommand{\hcr}{h_{\times}}

\newcommand{\hMN}{h^{\mu\nu}}
\newcommand{\hRS}{h^{\rho\sigma}}
\newcommand{\hMR}{h^{\mu\rho}}
\newcommand{\hRM}{h^{\rho\mu}}
\newcommand{\hRN}{h^{\rho\nu}}
\newcommand{\hNS}{h^{\nu\sigma}}
\newcommand{\hMS}{h^{\mu\sigma}}
\newcommand{\hNR}{h^{\nu\rho}}
\newcommand{\hAB}{h^{\alpha\beta}}
\newcommand{\hij}{h_{ij}}
\newcommand{\hIJ}{h^{ij}}
\newcommand{\hkl}{h_{kl}}
\newcommand{\hTTij}{h_{ij}^{\rm TT}}
\newcommand{\HTTij}{H_{ij}^{\rm TT}}
\newcommand{\dhTTij}{\dot{h}_{ij}^{\rm TT}}
\newcommand{\hTTab}{h_{ab}^{\rm TT}}

\newcommand{\thmn}{\tilde{h}_{\mu\nu}}
\newcommand{\thrs}{\tilde{h}_{\rho\sigma}}
\newcommand{\thmr}{\tilde{h}_{\mu\rho}}
\newcommand{\thns}{\tilde{h}_{\nu\sigma}}
\newcommand{\thms}{\tilde{h}_{\mu\sigma}}
\newcommand{\thnr}{\tilde{h}_{\nu\rho}}
\newcommand{\thrn}{\tilde{h}_{\rho\nu}}
\newcommand{\thab}{\tilde{h}_{\alpha\beta}}
\newcommand{\thMN}{\tilde{h}^{\mu\nu}}
\newcommand{\thRS}{\tilde{h}^{\rho\sigma}}
\newcommand{\thMR}{\tilde{h}^{\mu\rho}}
\newcommand{\thRM}{\tilde{h}^{\rho\mu}}
\newcommand{\thRN}{\tilde{h}^{\rho\nu}}
\newcommand{\thNS}{\tilde{h}^{\nu\sigma}}
\newcommand{\thMS}{\tilde{h}^{\mu\sigma}}
\newcommand{\thNR}{\tilde{h}^{\nu\rho}}
\newcommand{\thAB}{\tilde{h}^{\alpha\beta}}

\newcommand{\vvarphi}{\hat{\varphi}}
\newcommand{\hhmn}{\hat{h}_{\mu\nu}}
\newcommand{\hhrs}{\hat{h}_{\rho\sigma}}
\newcommand{\hhmr}{\hat{h}_{\mu\rho}}
\newcommand{\hhns}{\hat{h}_{\nu\sigma}}
\newcommand{\hhms}{\hat{h}_{\mu\sigma}}
\newcommand{\hhnr}{\hat{h}_{\nu\rho}}
\newcommand{\hhra}{\hat{h}_{\rho\alpha}}

\newcommand{\hhMN}{\hat{h}^{\mu\nu}}
\newcommand{\hhRS}{\hat{h}^{\rho\sigma}}
\newcommand{\hhMR}{\hat{h}^{\mu\rho}}
\newcommand{\hhNS}{\hat{h}^{\nu\sigma}}
\newcommand{\hhMS}{\hat{h}^{\mu\sigma}}
\newcommand{\hhNR}{\hat{h}^{\nu\rho}}
\newcommand{\hhAB}{\hat{h}^{\alpha\beta}}

\newcommand{\sh}{\mathsf{h}}
\newcommand{\shmn}{\mathsf{h}_{\mu\nu}}
\newcommand{\shrs}{\mathsf{h}_{\rho\sigma}}
\newcommand{\shmr}{\mathsf{h}_{\mu\rho}}
\newcommand{\shns}{\mathsf{h}_{\nu\sigma}}
\newcommand{\shms}{\mathsf{h}_{\mu\sigma}}
\newcommand{\shnr}{\mathsf{h}_{\nu\rho}}
\newcommand{\shra}{\mathsf{h}_{\rho\alpha}}
\newcommand{\shsb}{\mathsf{h}_{\sigma\beta}}
\newcommand{\shma}{\mathsf{h}_{\mu\alpha}}
\newcommand{\shna}{\mathsf{h}_{\nu\alpha}}
\newcommand{\shmb}{\mathsf{h}_{\mu\beta}}
\newcommand{\shas}{\mathsf{h}_{\alpha\sigma}}
\newcommand{\shab}{\mathsf{h}_{\alpha\beta}}
\newcommand{\shnb}{\mathsf{h}_{\nu\beta}}
\newcommand{\shcr}{\mathsf{h}_{\times}}
\newcommand{\shMN}{\mathsf{h}^{\mu\nu}}
\newcommand{\shRS}{\mathsf{h}^{\rho\sigma}}
\newcommand{\shMR}{\mathsf{h}^{\mu\rho}}
\newcommand{\shNS}{\mathsf{h}^{\nu\sigma}}
\newcommand{\shMS}{\mathsf{h}^{\mu\sigma}}
\newcommand{\shNR}{\mathsf{h}^{\nu\rho}}
\newcommand{\shAB}{\mathsf{h}^{\alpha\beta}}
\newcommand{\shij}{\mathsf{h}_{ij}}
\newcommand{\shIJ}{\mathsf{h}^{ij}}
\newcommand{\shkl}{\mathsf{h}_{kl}}
\newcommand{\shTTij}{\mathsf{h}_{ij}^{\rm TT}}
\newcommand{\shTTab}{\mathsf{h}_{ab}^{\rm TT}}

\newcommand{\bhmn}{\bar{h}_{\mu\nu}}
\newcommand{\bhrs}{\bar{h}_{\rho\sigma}}
\newcommand{\bhmr}{\bar{h}_{\mu\rho}}
\newcommand{\bhns}{\bar{h}_{\nu\sigma}}
\newcommand{\bhms}{\bar{h}_{\mu\sigma}}
\newcommand{\bhnr}{\bar{h}_{\nu\rho}}
\newcommand{\bhRS}{\bar{h}^{\rho\sigma}}
\newcommand{\bhMN}{\bar{h}^{\mu\nu}}
\newcommand{\bhNR}{\bar{h}^{\nu\rho}}
\newcommand{\bhMR}{\bar{h}^{\mu\rho}}
\newcommand{\bhAB}{\bar{h}^{\alpha\beta}}

\newcommand{\hax}{h^{\rm ax}}
\newcommand{\haxmn}{h^{\rm ax}_{\mu\nu}}
\newcommand{\hpol}{h^{\rm pol}}
\newcommand{\hpolmn}{h^{\rm pol}_{\mu\nu}}

\newcommand{\dgzz}{{^{(2)}g_{00}}}
\newcommand{\qgzz}{{^{(4)}g_{00}}}
\newcommand{\tgzi}{{^{(3)}g_{0i}}}
\newcommand{\dgij}{{^{(2)}g_{ij}}}
\newcommand{\zTzz}{{^{(0)}T^{00}}}
\newcommand{\dTzz}{{^{(2)}T^{00}}}
\newcommand{\dTii}{{^{(2)}T^{ii}}}
\newcommand{\uTzi}{{^{(1)}T^{0i}}}

\newcommand{\xm}{x^{\mu}}
\newcommand{\xn}{x^{\nu}}
\newcommand{\xr}{x^{\rho}}
\newcommand{\xs}{x^{\sigma}}
\newcommand{\xa}{x^{\a}}
\newcommand{\xb}{x^{\b}}

\newcommand{\hatk}{\hat{\bf k}}
\newcommand{\hatn}{\hat{\bf n}}
\newcommand{\hatx}{\hat{\bf x}}
\newcommand{\haty}{\hat{\bf y}}
\newcommand{\hatz}{\hat{\bf z}}
\newcommand{\hatr}{\hat{\bf r}}
\newcommand{\hatu}{\hat{\bf u}}
\newcommand{\hatv}{\hat{\bf v}}
\newcommand{\xim}{\xi_{\mu}}
\newcommand{\xin}{\xi_{\nu}}
\newcommand{\xia}{\xi_{\a}}
\newcommand{\xib}{\xi_{\b}}
\newcommand{\xiM}{\xi^{\mu}}
\newcommand{\xiN}{\xi^{\nu}}

\newcommand{\tA}{\tilde{\bf A} ({\bf k})}

\newcommand{\pam}{\pa_{\mu}}
\newcommand{\pal}{\pa_{\mu}}
\newcommand{\pan}{\pa_{\nu}}
\newcommand{\parho}{\pa_{\rho}}
\newcommand{\pas}{\pa_{\sigma}}
\newcommand{\paM}{\pa^{\mu}}
\newcommand{\paN}{\pa^{\nu}}
\newcommand{\paR}{\pa^{\rho}}
\newcommand{\paS}{\pa^{\sigma}}
\newcommand{\paa}{\pa_{\alpha}}
\newcommand{\pab}{\pa_{\beta}}
\newcommand{\pat}{\pa_{\theta}}
\newcommand{\paf}{\pa_{\phi}}

\newcommand{\Dam}{D_{\mu}}
\newcommand{\Dan}{D_{\nu}}
\newcommand{\Dar}{D_{\rho}}
\newcommand{\Das}{D_{\sigma}}
\newcommand{\DaM}{D^{\mu}}
\newcommand{\DaN}{D^{\nu}}
\newcommand{\DaR}{D^{\rho}}
\newcommand{\DaS}{D^{\sigma}}
\newcommand{\Daa}{D_{\alpha}}
\newcommand{\Dab}{D_{\beta}}

\newcommand{\DBm}{\bar{D}_{\mu}}
\newcommand{\DBn}{\bar{D}_{\nu}}
\newcommand{\DBr}{\bar{D}_{\rho}}
\newcommand{\DBs}{\bar{D}_{\sigma}}
\newcommand{\DBt}{\bar{D}_{\tau}}
\newcommand{\DBa}{\bar{D}_{\alpha}}
\newcommand{\DBb}{\bar{D}_{\beta}}
\newcommand{\DBM}{\bar{D}^{\mu}}
\newcommand{\DBN}{\bar{D}^{\nu}}
\newcommand{\DBR}{\bar{D}^{\rho}}
\newcommand{\DBS}{\bar{D}^{\sigma}}
\newcommand{\DBA}{\bar{D}^{\alpha}}

\newcommand{\GMnr}{{\Gamma}^{\mu}_{\nu\rho}}
\newcommand{\Glmn}{{\Gamma}^{\lambda}_{\mu\nu}}
\newcommand{\barGMnr}{{\bar{\Gamma}}^{\mu}_{\nu\rho}}
\newcommand{\GMns}{{\Gamma}^{\mu}_{\nu\sigma}}
\newcommand{\GInr}{{\Gamma}^{i}_{\nu\rho}}
\newcommand{\Rmn}{R_{\mu\nu}}
\newcommand{\Gmn}{G_{\mu\nu}}
\newcommand{\RMN}{R^{\mu\nu}}
\newcommand{\GMN}{G^{\mu\nu}}
\newcommand{\Rmnrs}{R_{\mu\nu\rho\sigma}}
\newcommand{\RMnrs}{{R^{\mu}}_{\nu\rho\sigma}}
\newcommand{\Tmn}{T_{\mu\nu}}
\newcommand{\Smn}{S_{\mu\nu}}
\newcommand{\Tab}{T_{\a\b}}
\newcommand{\TMN}{T^{\mu\nu}}
\newcommand{\TAB}{T^{\a\b}}
\newcommand{\TBmn}{\bar{T}_{\mu\nu}}
\newcommand{\TBMN}{\bar{T}^{\mu\nu}}
\newcommand{\TRS}{T^{\rho\sigma}}
\newcommand{\tmn}{t_{\mu\nu}}
\newcommand{\tMN}{t^{\mu\nu}}
\newcommand{\RUmn}{R_{\mu\nu}^{(1)}}
\newcommand{\RDmn}{R_{\mu\nu}^{(2)}}
\newcommand{\RTmn}{R_{\mu\nu}^{(3)}}
\newcommand{\RBmn}{\bar{R}_{\mu\nu}}
\newcommand{\RBmr}{\bar{R}_{\mu\rho}}
\newcommand{\RBnr}{\bar{R}_{\nu\rho}}

\newcommand{\dddM}{\kern 0.2em \raise 1.9ex\hbox{$...$}\kern -1.0em \hbox{$M$}}
\newcommand{\dddQ}{\kern 0.2em \raise 1.9ex\hbox{$...$}\kern -1.0em \hbox{$Q$}}
\newcommand{\dddI}{\kern 0.2em \raise 1.9ex\hbox{$...$}\kern -1.0em\hbox{$I$}}
\newcommand{\dddJ}{\kern 0.2em \raise 1.9ex\hbox{$...$}\kern-1.0em
\hbox{$J$}}
\newcommand{\dddcalJ}{\kern 0.2em \raise 1.9ex\hbox{$...$}\kern-1.0em
\hbox{${\cal J}$}}

\newcommand{\dddO}{\kern 0.2em \raise 1.9ex\hbox{$...$}\kern -1.0em
\hbox{${\cal O}$}}
\def\dddz{\raise 1.5ex\hbox{$...$}\kern -0.8em \hbox{$z$}}
\def\dddd{\raise 1.8ex\hbox{$...$}\kern -0.8em \hbox{$d$}}
\def\dddbd{\raise 1.8ex\hbox{$...$}\kern -0.8em \hbox{${\bf d}$}}
\def\ddbd{\raise 1.8ex\hbox{$..$}\kern -0.8em \hbox{${\bf d}$}}
\def\dddx{\raise 1.6ex\hbox{$...$}\kern -0.8em \hbox{$x$}}

\newcommand{\hti}{\tilde{h}}
\newcommand{\hf}{\tilde{h}_{ab}(f)}
\newcommand{\Hti}{\tilde{H}}
\newcommand{\fmin}{f_{\rm min}}
\newcommand{\fmax}{f_{\rm max}}
\newcommand{\frot}{f_{\rm rot}}
\newcommand{\fpol}{f_{\rm pole}}
\newcommand{\omax}{\o_{\rm max}}
\newcommand{\orot}{\o_{\rm rot}}
\newcommand{\op}{\o_{\rm p}}
\newcommand{\tmax}{t_{\rm max}}
\newcommand{\tobs}{t_{\rm obs}}
\newcommand{\fobs}{f_{\rm obs}}
\newcommand{\temis}{t_{\rm emis}}
\newcommand{\DE}{\D E_{\rm rad}}
\newcommand{\DEm}{\D E_{\rm min}}
\newcommand{\msun}{~{\rm M}_{\odot}}
\newcommand{\mtot}{M_{\rm tot}}
\newcommand{\rsun}{R_{\odot}}
\newcommand{\ogw}{\omega_{\rm gw}}
\newcommand{\fgw}{f_{\rm gw}}
\newcommand{\oL}{\omega_{\rm L}}
\newcommand{\kL}{k_{\rm L}}
\newcommand{\lL}{\l_{\rm L}}
\newcommand{\mns}{M_{\rm NS}}
\newcommand{\rns}{R_{\rm NS}}
\newcommand{\tret}{t_{\rm ret}}
\newcommand{\Sch}{Schwarzschild }
\newcommand{\rtid}{r_{\rm tidal}}

\newcommand{\ot}{\o_{\rm t}}
\newcommand{\mt}{m_{\rm t}}
\newcommand{\gt}{\g_{\rm t}}
\newcommand{\xit}{\tilde{\xi}}
\newcommand{\xtr}{\xi_{\rm t}}
\newcommand{\xtj}{\xi_{{\rm t},j}}
\newcommand{\dxtj}{\dot{\xi}_{{\rm t},j}}
\newcommand{\ddxtj}{\ddot{\xi}_{{\rm t},j}}
\newcommand{\teff}{T_{\rm eff}}
\newcommand{\samp}{S_{\xi_{\rm t}}^{\rm ampl}}

\newcommand{\mpl}{M_{\rm Pl}}
\newcommand{\mplr}{m_{\rm Pl}}
\newcommand{\mgut}{M_{\rm GUT}}
\newcommand{\lpl}{l_{\rm Pl}}
\newcommand{\tpl}{t_{\rm Pl}}
\newcommand{\ls}{\lambda_{\rm s}}
\newcommand{\Ogw}{\Omega_{\rm gw}}
\newcommand{\hogw}{h_0^2\Omega_{\rm gw}}
\newcommand{\hn}{h_n(f)}

\newcommand{\sinc}{{\rm sinc}\, }
\newcommand{\Ein}{E_{\rm in}}
\newcommand{\Eout}{E_{\rm out}}
\newcommand{\Et}{E_{\rm t}}
\newcommand{\Er}{E_{\rm refl}}
\newcommand{\lm}{\l_{\rm mod}}

\newcommand{\mrI}{\mathrm{I}}
\newcommand{\mrJ}{\mathrm{J}}
\newcommand{\mrW}{\mathrm{W}}
\newcommand{\mrX}{\mathrm{X}}
\newcommand{\mrY}{\mathrm{Y}}
\newcommand{\mrZ}{\mathrm{Z}}
\newcommand{\mrM}{\mathrm{M}}
\newcommand{\mrS}{\mathrm{S}}

\newcommand{\rmI}{\mathrm{I}}
\newcommand{\rmJ}{\mathrm{J}}
\newcommand{\rmW}{\mathrm{W}}
\newcommand{\rmX}{\mathrm{X}}
\newcommand{\rmY}{\mathrm{Y}}
\newcommand{\rmZ}{\mathrm{Z}}
\newcommand{\rmM}{\mathrm{M}}
\newcommand{\rmS}{\mathrm{S}}
\newcommand{\rmU}{\mathrm{U}}
\newcommand{\rmV}{\mathrm{V}}


\newcommand{\et}{{{\bf e}^t}}
\newcommand{\etm}{{\bf e}^t_{\mu}}
\newcommand{\etn}{{\bf e}^t_{\nu}}

\newcommand{\er}{{{\bf e}^r}}
\newcommand{\erm}{{\bf e}^r_{\mu}}
\newcommand{\ern}{{\bf e}^r_{\nu}}

\newcommand{\hz}{H^{(0)}}
\newcommand{\hu}{H^{(1)}}
\newcommand{\hd}{H^{(2)}}
\newcommand{\thz}{\tilde{H}^{(0)}}
\newcommand{\thu}{\tilde{H}^{(1)}}
\newcommand{\thd}{\tilde{H}^{(2)}}
\newcommand{\tK}{\tilde{K}}
\newcommand{\tZ}{\tilde{Z}}
\newcommand{\tQ}{\tilde{Q}}

\newcommand{\inT}{\int_{-\infty}^{\infty}}
\newcommand{\intz}{\int_{0}^{\infty}}
\newcommand{\Dl}{\int{\cal D}\lambda}

\newcommand{\fnl}{f_{\rm NL}}

\newcommand{\ode}{\Omega_{\rm DE}}
\newcommand{\oma}{\Omega_{M}}
\newcommand{\ora}{\Omega_{R}}
\newcommand{\ovac}{\Omega_{\rm vac}}
\newcommand{\ola}{\Omega_{\Lambda}}
\newcommand{\oxi}{\Omega_{\xi}}
\newcommand{\oga}{\Omega_{\gamma}}

\newcommand{\lc}{\Lambda_c}
\newcommand{\rde}{\rho_{\rm DE}}
\newcommand{\wde}{w_{\rm DE}}
\newcommand{\rvac}{\rho_{\rm vac}}
\newcommand{\rlam}{\rho_{\Lambda}}

\newcommand{\NM}[1]{{\color{blue}{[\bf{NM}}: #1]}}

\maketitle
\flushbottom

\section{Introduction}

The detection of gravitational waves (GWs) in the last few years has opened a new  way
of observing the Universe~~\cite{LIGOScientific:2016aoc,LIGOScientific:2017vwq,LIGOScientific:2017zic,LIGOScientific:2017ync,LIGOScientific:2020ibl,KAGRA:2021vkt,KAGRA:2021duu,LIGOScientific:2021sio,LIGOScientific:2021aug}. A new generation  of ground-based GW detectors,   referred as third-generation (3G) or next-generation GW detectors,  is under study, to overcome the limitations imposed by current detector infrastructures. The European project is Einstein Telescope (ET)  \cite{Punturo:2010zz,Hild:2010id,Maggiore:2019uih}, while the US community effort is represented by the Cosmic Explorer (CE) project~\cite{Reitze:2019iox,Evans:2021gyd,Evans:2023euw}.

Einstein Telescope  is meant to be a flagship for European science in the 2030s and it is therefore important, in the current stage of development, to investigate different options for its configuration, also within different global detector networks.
Several studies have been conducted in the last few years, to compare  the scientific potential of ET as well as of various 3G  detector networks  \cite{Borhanian:2022czq,Branchesi:2023mws}; see also refs.~\cite{Puecher:2023twf,Bhagwat:2023jwv,Franciolini:2023opt,Iacovelli:2023nbv} for follow-up studies elaborating on various aspects of ref.~\cite{Branchesi:2023mws}, refs.~\cite{Gupta:2023lga,Iacovelli:2024mjy} for   studies of the capabilities of further 3G network configurations, and ref.~\cite{Abac:2025saz} for a recent comprehensive study of the ET Science case, also in various network configurations. In particular, in ref.~\cite{Branchesi:2023mws} two different  geometries for ET were compared, a single-site triangular geometry made of 3 nested detectors (with each detector made by a high-frequency interferometer and a cryogenic low-frequency interferometer), located underground (that we will denote as ET-$\Delta$), and a network made of two identical  L-shaped detectors (`2L' in the following), located in two different sites within Europe, but still underground, and  both made by a high-frequency interferometer and a cryogenic  low-frequency interferometer.
These different configurations for ET were considered in a ET-only scenario, as well as  
in a broader world-wide network including also  a single 40-km Cosmic Explorer (CE) detector, or  two CE detectors with arm-lengths of 20 and 40~km, respectively.

Recently, a report of the NSF MPS AC Subcommittee
on Next-Generation Gravitational-Wave Detector Concepts~\cite{Evans:2023euw} included a network of a single CE-40km together with ET (which was considered in the triangle configuration)  in a list of  recommended  world-wide next-generation networks.\footnote{Note, however, that the baseline configuration for the CE project is a pair of detectors, one with 20 km arms and the other with 40 km arms; see 
\cite{Gupta:2023lga} for an exhaustive study of the different CE configurations, with and without ET.}
The performance of such a network, as well as of a network made by a single CE-40km together with ET in its 2L configuration, was already discussed in detail in \cite{Branchesi:2023mws}.
In this paper, we extend the analysis to the situation in which the European detector is made of just a single L-shaped underground detector, with the amplitude spectral density (ASD) of ET,\footnote{The ASD, $S^{1/2}_n(f)$, with dimensions ${\rm Hz}^{-1/2}$, is the square root of the noise spectral density $S_n(f)$, defined by
$\langle \tilde{n}^*(f) \tilde{n}(f')\rangle=(1/2) S_n(f) \delta(f-f')$, where $\tilde{n}(f)$ is the Fourier transform of the noise, $n(t)$;
see  section~7.1 of \cite{Maggiore:2007ulw} for definitions and conventions.} and 15~km or 20~km arm-length.
Such a network should  be  seen as an intermediate step toward the realization  of a larger world-wide detector network, featuring a 2L ET  configuration together with a 40-km CE detector.

Of course, the different detector options that we consider here  have different levels of complexity. The ET-$\Delta$ configuration involves 3 nested detectors of 10 km arms,  ET-2L involves two detectors with 15 km arms, and the single-L just one  detector, of 15-km or 20~km arms (where, in all cases, a detector is actually made by two interferometers, a low-frequency and a high-frequency interferometer, so ET-$\Delta$ is made of 6 interferometers, ET-2L of 4, and the single L of 2).\footnote{We stick to the recommended terminology from the 2020 ET Design Report Update,
\url{https://apps.et-gw.eu/tds/?r=18715}: the high-frequency (HF) and low-frequency  (LF) interferometers that make the so-called ``xylophone'' configuration are indeed referred to as ``interferometers''. The combination of a HF interferometer  and a LF interferometer (whether in a L-shaped geometry, or with arms at $60^{\circ}$ as in the triangle configuration)
is called a ``detector''.  The whole ensemble of detectors is called an ``observatory''.} We already know from the study in 
\cite{Branchesi:2023mws} that, for coalescing binaries, the ET-2L-15km configuration performs better than the 10-km triangle when they are both considered in isolation, and that the differences become less evident when they are put in a network with a 40-km CE. 
Obviously, the configuration with a single L is bound to be less
performant than the 2L configuration or the triangle, but our aim is to understand how much we lose in term of science,
with the single-L detector, for example in case of delay in the implementation of the full European ET network or in case of prolonged absence of one of the two ET nodes, in a scenario that also features a 40-km CE. As discussed in section~8.1.3 of \cite{Branchesi:2023mws}, a single L-shaped detector with the ET sensitivity, not inserted in a global 3G detector network,  is not capable of delivering the science expected from the next generation of GW detectors. Considering that CE has a different sensitivity curve, and in particular does not have access to the low frequencies accessible to ET, it is therefore a non-trivial question whether a network made by a single L in Europe with the ET sensitivity curve and a 40~km CE  in the US  has a valid science case. This is the main question that we address in this paper.

In the context of the ground-based GW detectors that might be operating at the time of ET, it is also quite interesting to study the effect of adding  to the  detector network also LIGO-India, which is currently under construction in Aundha, India. LIGO-India will have 4~km arms, 
and is expected to be operational in the 2030s at A+ sensitivity \cite{KAGRA:2013rdx}, with upgrades
at A$^{\#}$ sensitivity levels \cite{Fritschel:2022}.
Indeed,   recent works that studied  a CE-40km+ LIGO-India network \cite{Pandey:2024mlo} found that, for some observables such as sky localization, the shorter arm-length of LIGO-India (compared to ET or to CE) is compensated by the long baselines between the detectors, leading to quite competitive results. This was confirmed by a study of CE-40km+ LIGO-India +ET, with ET in the triangle or 2L configurations, performed in \cite{Abac:2025saz}.
We therefore add to the list of networks that we study a network featuring a single L-shaped underground detector with the  ASD of ET, with  15~km  arms,
together with a single 40-km CE located in the US, and  LIGO-India (that we take at  A$^\#$ sensitivity).
Again, we see this as a possible intermediate step toward a broader network where ET is in its 2L configuration.

Finally, again in the spirit of examining the performance of networks that could  represent intermediate steps toward the full planned detector sensitivities and detector networks, one should take into account that the ET design involves two interferometers (LF and HF) for each detector, one optimized for low frequencies and one for high frequencies.  The LF instrument is expected to be  more difficult to commission. It is therefore interesting to consider also an intermediate step in which, in ET, only the HF instrument is operative, as was already done in \cite{Branchesi:2023mws}. We will therefore extend our analysis to the situation in which the network is composed by CE-40km in the US, together with a single L in Europe with the sensitivity of ET with the HF-only instrument.

\section{Network configurations}

\begin{figure}[tbp]
    \centering
    \includegraphics[width=.6\textwidth]{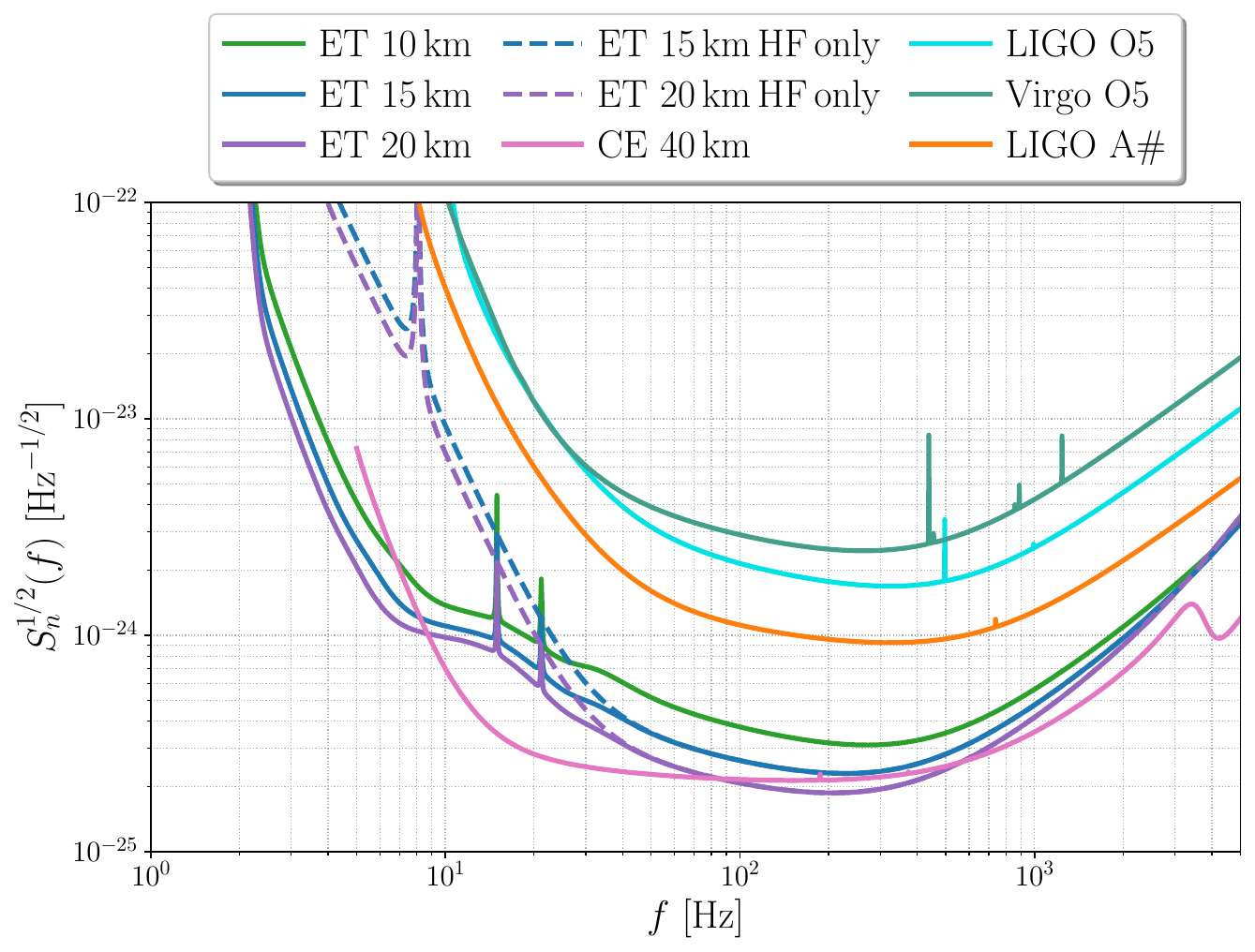}
    \caption{Amplitude spectral densities (ASDs) used in this work. Note that the ASDs are always defined as if we had a single L-shaped detector of the given arm-length; for the triangle, they are then combined as discussed in \cite{Branchesi:2023mws}.}
    \label{fig:All_ASDs}
\end{figure}
The ASD of the  detectors that we are considering are shown in 
\autoref{fig:All_ASDs}.\footnote{For ET we use the same ASD as in \cite{Branchesi:2023mws}, which is the most recent publicly available ASD. It should be kept in mind, however, that at the current stage of development, when the design of ET has not yet been frozen, the sensitivity curve  is  still evolving.}$^{,}$\footnote{Note that
the ASDs shown in the figure always refer to a single L-shaped detector. As discussed in section~2 of \cite{Branchesi:2023mws},
to obtain the sensitivity of ET in the triangle configuration 
one must take into account that the triangle is made of three nested detectors  with an opening angle of $60^{\circ}$. 
One must then  project the GW tensor of the incoming wave onto each of these three  components [see e.g. eqs.~(9)--(11) of \cite{Jaranowski:1998qm} for explicit expressions], and 
combine the results at the level of the SNR  and parameter estimation.}
We  consider the following network configurations:

\begin{itemize}

\item ET in the 10-km triangle configuration (ET-$\Delta$), with its standard design, therefore underground and in a xylophone configuration involving the LF and HF instruments, together with a single 40-km CE located in the US. We denote this network as (ET-$\Delta$ + CE-40km).

\item ET in the configuration of two underground  L-shaped detectors with 15 km arms, together with a single 40-km CE located in the US. 
We  denote this network as (ET-2L + CE-40km). 

\item A single L-shaped underground detector with the  ASD of ET, with  either 15~km  or 20~km arms,
together with a single 40-km CE located in the US.
We denote these networks  as (1${\rm L}_{\rm EU}^{15\, \rm km}$ + CE-40km) and 
(1${\rm L}_{\rm EU}^{20\, \rm km}$ + CE-40km), respectively.

\item A single L-shaped underground detector with the  ASD of ET, with  15~km  arms,
together with a single 40-km CE located in the US and  LIGO-India (that we take at  A$^\#$ sensitivity). We denote this network as   
(1${\rm L}_{\rm EU}^{15\, \rm km}$ + CE-40km + LIGO-I).\footnote{Note that we purposely refrain from using the label ``ET" for the configurations where the European detectors is a single L. The label ET is a ``trademark" that must be reserved only to the  currently officially recognized ET configurations, i.e.  triangle or 2L.} 

\end{itemize}

Concerning the locations, we place  for definiteness the triangle in the Sardinia candidate site, but essentially no significant difference would be obtained putting it in the Meuse-Rhine (EMR) candidate site. In the ET-2L configuration we put the two L-shaped detectors in the Sardinia and EMR candidate sites; basically the same results would be obtained using, as locations for the two L-shaped detectors,  the Sardinia candidate site together  the recently proposed candidate site in Saxony, since the cord distances are very similar, see footnote~52 in \cite{Branchesi:2023mws}. In the 1L case, we put the single L  in the Sardinia candidate site, but again, at the level of our analysis,  no significant difference would be obtained putting it in the EMR or Saxony  candidate sites. For the 40-km CE in the US,  we use for definiteness  the location and orientation in Table III of \cite{Borhanian:2020ypi}.
The results for (ET-$\Delta$ + CE-40km) and  
(ET-2L + CE-40km) were already shown in \cite{Branchesi:2023mws}, and here we will plot them together with the results for (1${\rm L}_{\rm EU}^{15\, \rm km}$ + CE-40km), (1${\rm L}_{\rm EU}^{20\, \rm km}$ + CE-40km) and (1${\rm L}_{\rm EU}^{15\, \rm km}$ + CE-40km + LIGO-I), which are new. Note that, in ref.~\cite{Branchesi:2023mws},  the ET-2L case was studied both when the two L-shaped detectors are  misaligned, with a relative angle close to $45^{\circ}$, and when they are  parallel. Here, in order not to overburden the plots, we will only show the results for (ET-2L + CE-40km) when ET-2L are in the misaligned configuration, which provides the best results for compact binary coalescences;  when we wish to emphasize this,   we will actually  denote this configuration as
(ET-2L-mis + CE-40km), otherwise, in the interest of readability, we will just use the notation (ET-2L + CE-40km). 
We will instead use a generic orientation with respect to the CE detector in the US (we use the same orientations as in~\cite{Branchesi:2023mws}).  
In the (1${\rm L}_{\rm EU}$ + CE-40km) configurations, we will set the two detector at $45^{\circ}$ with respect to the great circle joining them.

We will also compare  with the results that can be obtained with the ET-$\Delta$ configuration not inserted in a network with CE, as well as with the results that 
could be obtained by the most advanced  2G detector network,  namely  LIGO Hanford, LIGO Livingston, Virgo, KAGRA and LIGO India,  using the publicly available best sensitivities that are planned to be achieved by the end of the O5 run \cite{KAGRA:2020rdx}. We denote this network as LVKI~O5.

Our methodology (astrophysical populations, parameter estimation codes, waveforms, etc.) is identical to the one  already followed in Section~3 of ref.~\cite{Branchesi:2023mws}, whom we refer the reader for more details. In particular, for compact binary coalescences, we perform parameter estimation in the Fisher matrix approximation,  using the 
\texttt{GWFAST} code~\cite{Iacovelli:2022bbs,Iacovelli:2022mbg}.\footnote{See also  \cite{Borhanian_2021,Dupletsa:2022scg,Chan:2018csa,Li:2021mbo,Pieroni:2022bbh,Begnoni:2025oyd} for other Fisher matrix codes tuned to 3G detectors; these codes have been successfully cross-checked in the context of the ET activities~\cite{Branchesi:2023mws}.}$^{,}$\footnote{We do not include the effect of correlated noise; recently, the effect of correlated noise on parameter estimation has been studied in
\cite{Wong:2024hes}. Although the results in \cite{Wong:2024hes} apply to two  colocated L-shaped detectors, and not to the triangle configuration, when using the realistic coherence values found in
\cite{Janssens:2024jln}, they point toward the fact that, for parameter estimation of coalescing binaries, correlated noise can be neglected. Correlated noise are, however, significant for stochastic backgrounds in the triangle configuration, as discussed in Section 5.4 of ref.~\cite{Branchesi:2023mws}, and as we will discuss again below.}

\section{Results}

In this section we present the results of the analysis. The structure of the plots and the definition of the  various observables are the same as in \cite{Iacovelli:2022bbs,Branchesi:2023mws,Iacovelli:2024mjy}.

\subsection{Horizons}

\begin{figure}[tb]
    \centering
    \includegraphics[height=.5\textwidth]{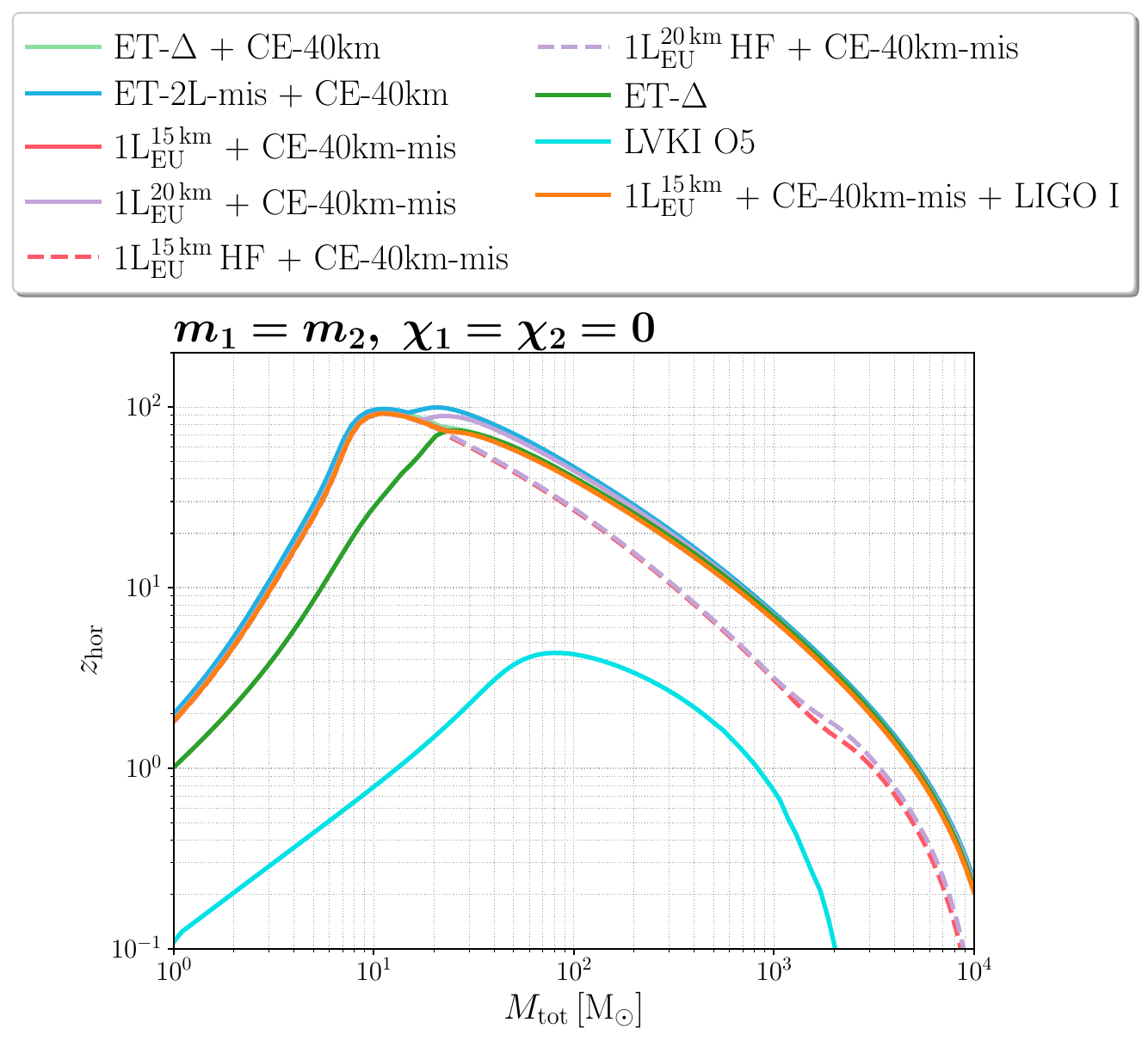}\\
    \caption{Detector horizons for equal mass, non-spinning binaries for the various detector geometries considered. We also add, for comparison, LVKI~O5.}
    \label{fig:All_horizons}
\end{figure}

\autoref{fig:All_horizons} shows the detection horizons    for equal-mass non-spinning coalescing binaries,  as a function of the total mass of the binary, for the  3G  networks considered
and, for comparison, also for ET-$\Delta$ and LVKI~O5.
We see  that, in terms of detection horizons,  all the  3G configurations  (with ET featuring the HF and LF interferometers) are quite similar, except for ET-$\Delta$ in isolation, which is less performant at low masses, and they all allow us to make a large jump with respect to the best possible 2G detector network.  In particular,  for $\mtot=2.7\msun$, a typical value for the total mass of a  BNS, 
the values of the horizons  are given in \autoref{tab:horizon_tabs}. We see that, in all cases, the horizons are well beyond the peak of the star formation rate, which is at a redshift $z_p\sim 2-3$, and basically cover the whole range of redhsifts relevant for BNS. We also observe that the loss of the LF instrument does not impact significantly the horizon distance for $M_{\rm tot}$ below about $20\msun$, but becomes quite significant above.

\begin{table}[tbp]
    \centering
    \begin{tabular}{!{\vrule width 1pt}l|c!{\vrule width 1pt}}
        \toprule
        \midrule
        \multicolumn{1}{!{\vrule width 1pt}c|}{Detector configuration} & $z_{\rm hor}(2.7\msun)$ \\
        \midrule
        ET-$\Delta$ + CE-40km & 8.5 \\
        ET-2L-mis + CE-40km & 9.0 \\
        ${\rm 1L}_{\rm EU}^{\rm 15\,km}$ + \textrm{CE-40km-mis} & 8.1 \\
        ${\rm 1L}_{\rm EU}^{\rm 20\,km}$ + \textrm{CE-40km-mis} & 8.3 \\
        ${\rm 1L}_{\rm EU}^{\rm 15\,km}$\,HF + \textrm{CE-40km-mis} & 8.0 \\
        ${\rm 1L}_{\rm EU}^{\rm 20\,km}$\,HF + \textrm{CE-40km-mis} & 8.0 \\
        ET-$\Delta$ & 3.3 \\
        ${\rm 1L}_{\rm EU}^{\rm 15\,km}$ + \textrm{CE-40km-mis + LIGO I} & 8.1 \\
        \midrule
        LVKI O5 & 0.3 \\
        \midrule
        \bottomrule
    \end{tabular}
    
    \caption{Horizon redshifts for equal mass non-spinning binaries evaluated at a source-frame total mass $M_{\rm tot} = 2.7\msun$ for the various configurations considered.}
    \label{tab:horizon_tabs}
\end{table}

\subsection{Parameter reconstruction for BBHs}\label{sect:PEBBH}

The results from the accuracy of parameter reconstruction for BBHs is shown in Fig.~\ref{fig:PE_BBH_wCE}. \autoref{tab:BBH_numbers_loc_wCE} gives, more quantitatively, the number of detections with some cuts on angular localization or on luminosity distance. Here and in the following, all the results are given for one year of data, with the duty cycle computed as in~\cite{Branchesi:2023mws}.

As expected, among the world-wide networks not involving LIGO-India, the hierarchy is that the best results come from (ET-2L-mis + CE-40km), followed by 
(ET-$\Delta$ + CE-40km), and then by 
(1${\rm L}_{\rm EU}$ + CE-40km) [with (1${\rm L}_{\rm EU}^{20\, \rm km}$ + CE-40km) only marginally better than
(1${\rm L}_{\rm EU}^{15\, \rm km}$ + CE-40km)], but
we see that the results are quite similar among these  3G global networks: the differences among the configurations are in general within about a factor of 2, and all these configurations provide a large jump compared to 2G detectors. As far as BBHs are concerned, the (1${\rm L}_{\rm EU}$ + CE-40km)
option therefore appears to maintain a relevant scientific output, also in the version with 15~km arms. 

It is also quite interesting to observe, from \autoref{tab:BBH_numbers_loc_wCE}, that for angular localization the network (1${\rm L}_{\rm EU}^{15\, \rm km}$ + CE-40km + LIGO-I)  is the second best, with about $3.2\times 10^3$ BBHs/yr localized to better than $1\, {\rm deg}^2$, to be compared with $3.7\times 10^3$ BBHs/yr 
for (ET-2L-mis  + CE-40km)  (which gives the best results) and
$2.4\times 10^3$ BBHs/yr for
(ET-$\Delta$ + CE-40km). This confirm that, for angular localization, the long baseline of a network involving LIGO-India compensates for its shorter arm-length \cite{Gupta:2023lga,Pandey:2024mlo}, providing a very interesting angular localization. 
For the accuracy on luminosity distance, considering for instance the events with $d_L$ measured to better than $1\%$, the best network is again
 (ET-2L-mis  + CE-40km), with $4.3\times 10^3$ BBH/yr, followed now by 
 (ET-$\Delta$ + CE-40km) with $2.9\times 10^3$ BBH/yr, while (1${\rm L}_{\rm EU}^{15\, \rm km}$-mis + CE-40km + LIGO-I) would get $2.1\times 10^3$ BBH/yr.

A similar conclusion emerges from \autoref{fig:histz_BBH_wCE}, which gives the redshift distribution of BBH golden events, i.e. events detected with especially high SNR, or especially good accuracy on luminosity distance, or especially good angular localization. We also observe that each of these international network give results significantly  better than ET-$\Delta$ in isolation, particularly for angular localization and luminosity distance where the improvements in the number of events with given accuracy can be of two orders of magnitude, see \autoref{tab:BBH_numbers_loc_wCE}. This is non-trivial considering that ET-$\Delta$  is made of six interferometers while, for instance, ET-2L + CE-40km only by three (the two interferometers in  ET-2L, HF and LF, plus the single interferometer of CE-40km).

\begin{figure}[!tbp]
    \centering
    \includegraphics[width=.95\textwidth]{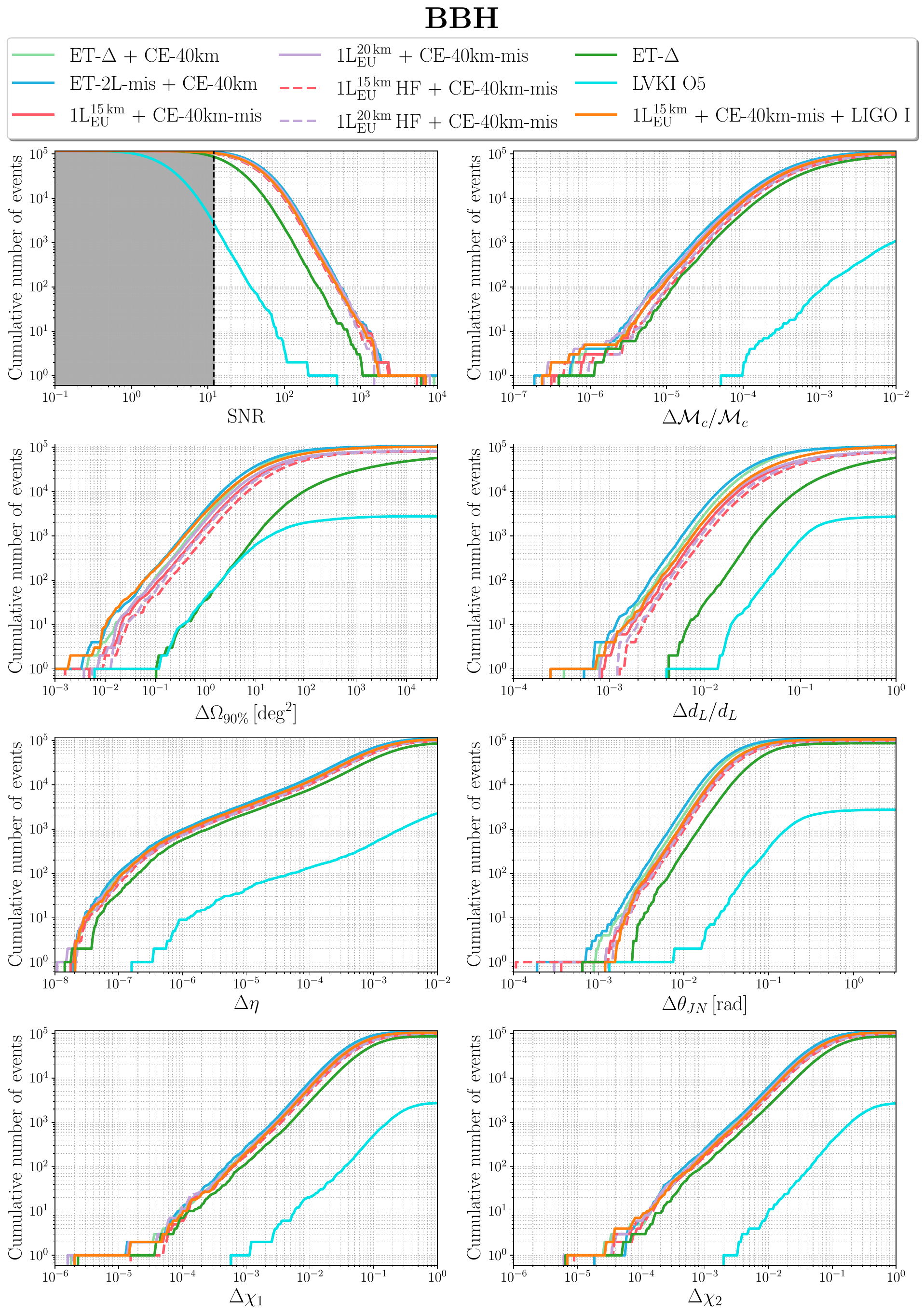}
    \caption{Cumulative distributions of the number of detections per year, for the SNRs and for the error on the parameters, for BBH signals, for the various detector configurations considered.}
    \label{fig:PE_BBH_wCE}
\end{figure}

\subsection{Parameter reconstruction for BNSs}\label{sect:PEBNS}

The corresponding results for BNSs are shown in \autoref{fig:PE_BNS_wCE} and \autoref{tab:BNS_numbers_loc_wCE}. Again, let us compare first the results for the networks not involving LIGO-India.
From \autoref{fig:PE_BNS_wCE} we see that, while the results for the SNR distribution and for the accuracy of reconstruction of the chirp mass ${\cal M}_c$, of the symmetric mass ratio $\eta$, and of the tidal deformability $\tilde{\Lambda}$ are  similar among the 3G world-wide networks considered (when ET is in the LF+HF configuration), a large difference appears in angular localization, luminosity distance, orbit inclination $\theta_{JN}$
and polarization angle $\psi$, where the (ET-2L-mis + CE-40km) and
(ET-$\Delta$ + CE-40km) configurations are better than 
(1${\rm L}_{\rm EU}$ + CE-40km)   by one order of magnitude.  This is due to the fact that for some parameters (such as masses and tidal deformability) the quality of parameter estimation depends mainly on the network SNR, and this does not change significantly between (ET-2L-mis + CE-40km) and  (1${\rm L}_{\rm EU}$ + CE-40km). In contrast, for angular localization triangulation is very important, so adding one more  well-separated detector to the network improves it significantly, inducing also a significant improvement in the parameters, such as luminosity distance or polarization angle, that are more sensitive  to the accuracy on angular localization.
The same effect is also shown, for some specific cuts on angular localization or on luminosity distance, in \autoref{tab:BNS_numbers_loc_wCE}. The corresponding results for ``golden events" are shown in \autoref{fig:histz_BNS_wCE}.
It is also interesting to observe that the (1${\rm L}_{\rm EU}$ + CE-40km)  configurations are, in turn, significantly better than the triangle in isolation, in particular for angular localization, luminosity distance and  polarization angle.

Adding LIGO-India to the (1${\rm L}_{\rm EU}$ + CE-40km)
network brings very interesting improvements. Now angular localization becomes comparable among  (ET-$\Delta$ + CE-40km), (ET-2L-mis + CE-40km) and 
(1${\rm L}_{\rm EU}^{15\, \rm km}$ + CE-40km + LIGO-I). 
For instance, for the BNS localized to better than $10\, {\rm deg}^2$, the best results are obtained as usual with (ET-2L-mis + CE-40km), with $3.8\times 10^3$ events/yr, followed by (1${\rm L}_{\rm EU}^{15\, \rm km}$ + CE-40km + LIGO-I) with $3.2\times 10^3$ events/yr and (ET-$\Delta$ + CE-40km), with $2.4\times 10^3$ events/yr, see \autoref{tab:BNS_numbers_loc_wCE}. All other configurations considered fall well behind; in particular, ET-$\Delta$ in isolation only gets 8~events/yr with this localization. For the accuracy on luminosity distance, considering for instance the events with an error on $d_L$ smaller than $5\%$, we see from \autoref{tab:BNS_numbers_loc_wCE} that 
the best results are obtained again by  (ET-2L-mis + CE-40km), with 1040 events/yr, followed now by
 (ET-$\Delta$ + CE-40km) with 535 events/yr and 
(1${\rm L}_{\rm EU}^{15\, \rm km}$ + CE-40km + LIGO-I) with 334  events/yr.

\begin{figure}[!tbp]
    \centering
    \begin{tabular}{c c c}
       \includegraphics[width=.32\textwidth]{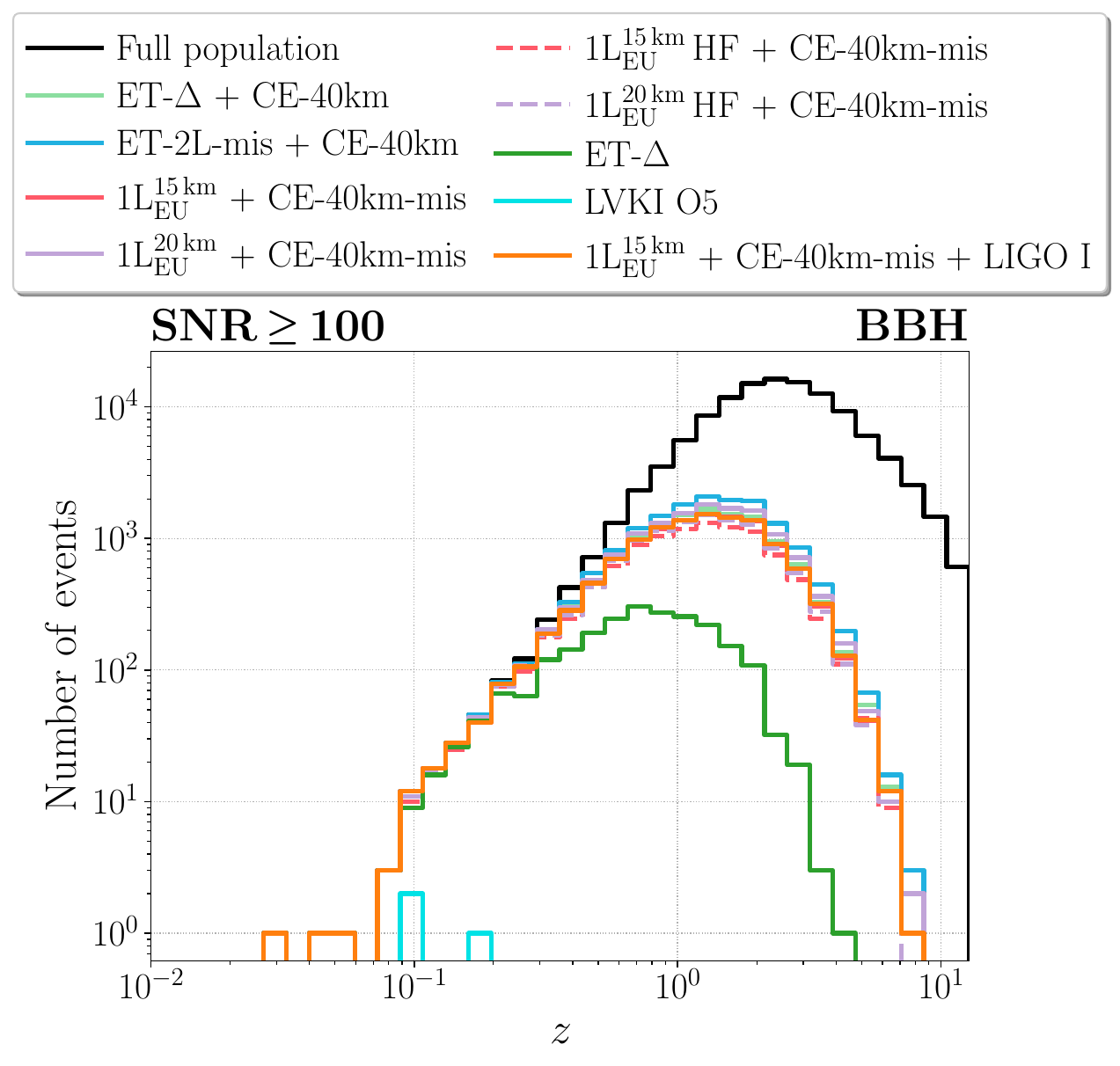}  & \includegraphics[width=.32\textwidth]{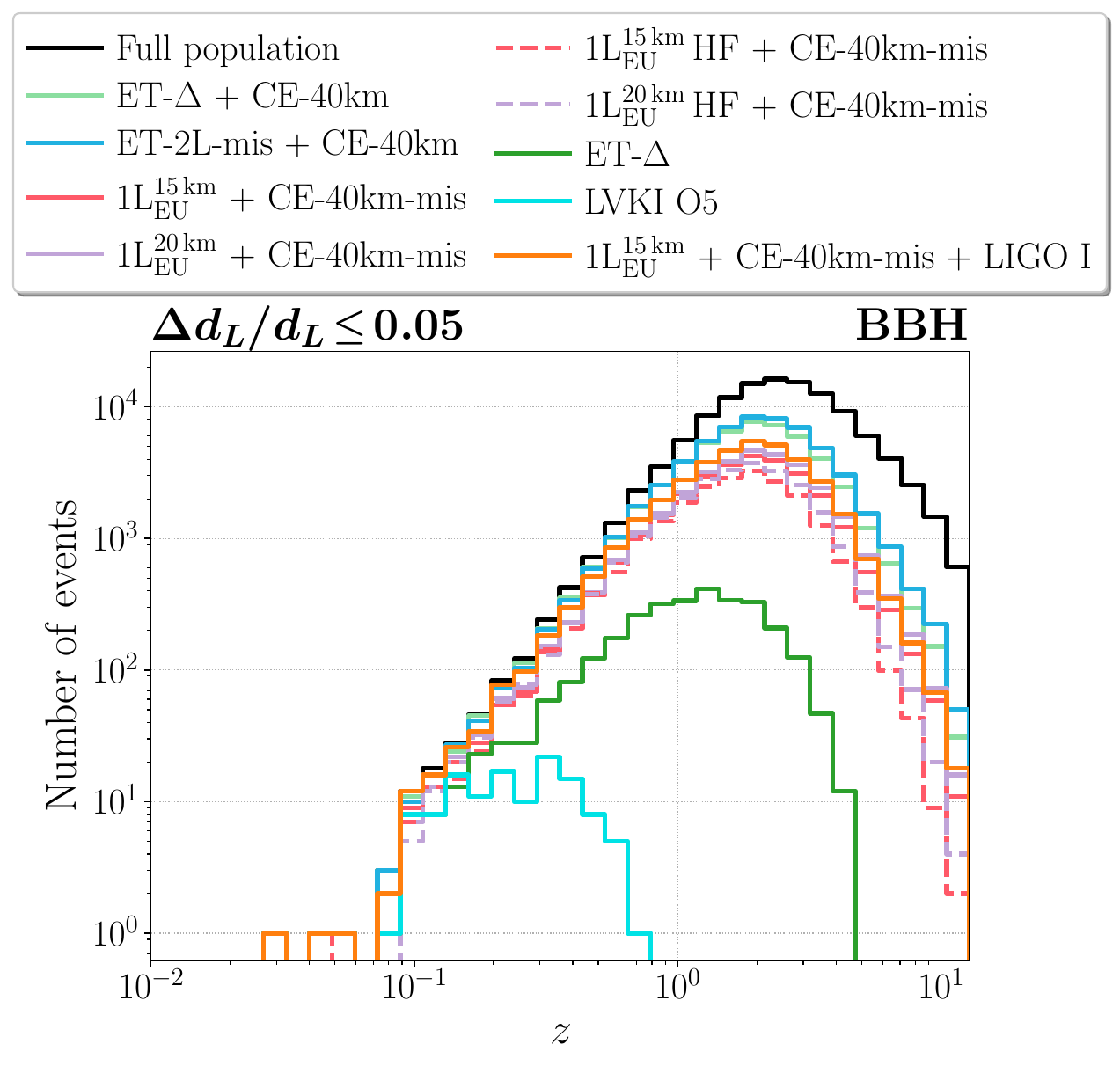} & \includegraphics[width=.32\textwidth]{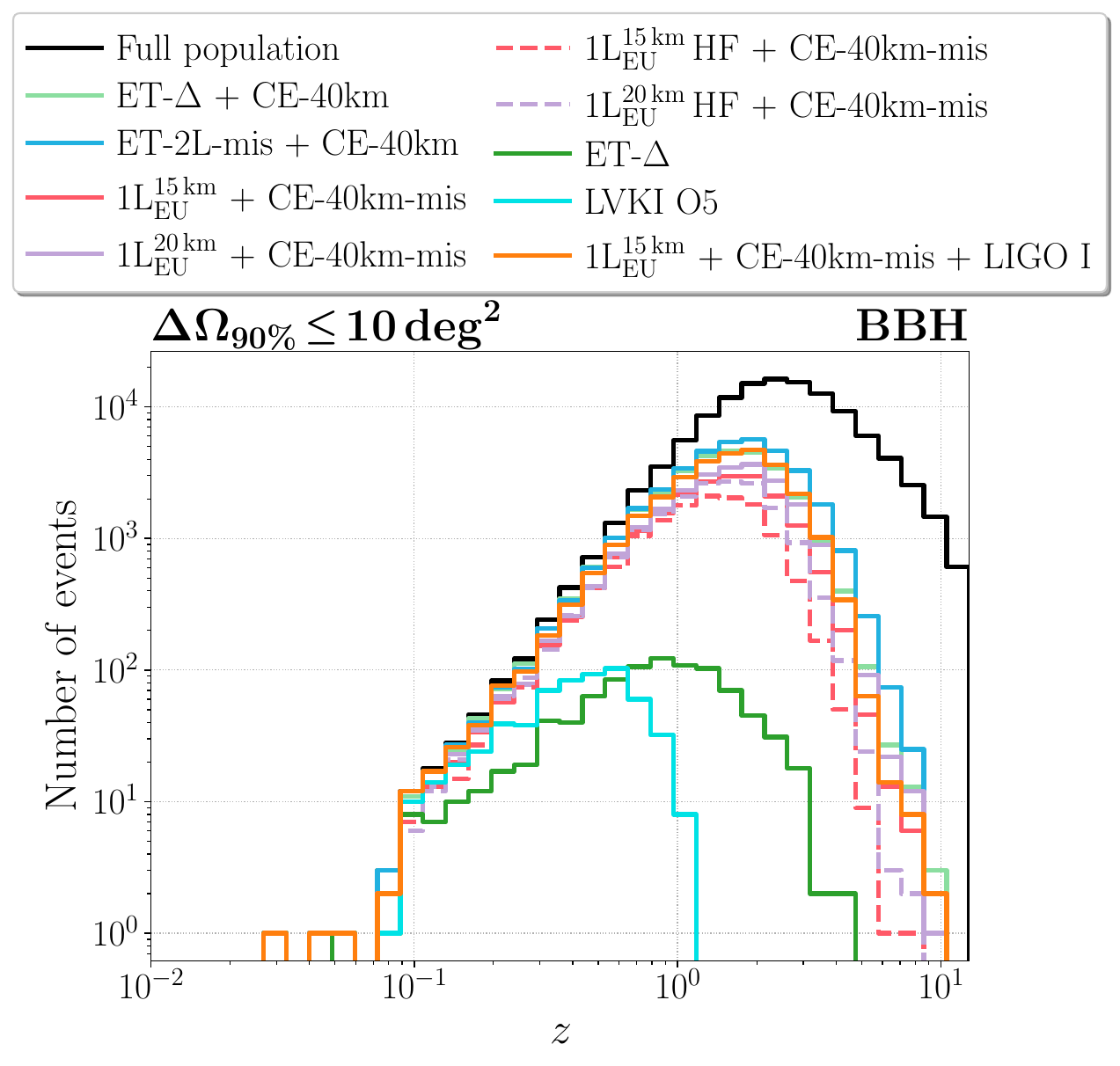} 
    \end{tabular}
    \caption{Redshift distribution of BBHs detected with ${\rm SNR} \geq 100$ (left column), or relative error on the luminosity distance $\Delta d_L/d_L \leq 0.05$ (central column), or sky location $\Delta\Omega_{90\%} \leq 10~{\rm deg}^2$ (right column) for the various detector configurations considered.}
    \label{fig:histz_BBH_wCE}
\end{figure}

\begin{table}[!tbp]
    \centering
    \begin{tabular}{!{\vrule width 1pt}m{5.8cm}|S[table-format=4.0]|S[table-format=5.0]|S[table-format=3.0]|S[table-format=4.0]!{\vrule width 1pt}}
        \toprule
        \midrule
        \multicolumn{5}{!{\vrule width 1pt}c!{\vrule width 1pt}}{\textbf{BBH}}\\
        \midrule
        \hfil\multirow{3}{*}{\shortstack[c]{Detector\\configuration}}\hfill & \multicolumn{4}{c!{\vrule width 1pt}}{Detections with} \\
        & \multicolumn{2}{c|}{$\Delta\Omega_{90\%}\leq$} & \multicolumn{2}{c!{\vrule width 1pt}}{$\Delta d_L/d_L \leq$}\\
        \cmidrule(lr){2-5}
        & \multicolumn{1}{c|}{$1~{\rm deg}^2$} & \multicolumn{1}{c|}{$10~{\rm deg}^2$}  & \multicolumn{1}{c|}{$5\times10^{-3}$} & \multicolumn{1}{c!{\vrule width 1pt}}{$10^{-2}$}\\
        \midrule
        \textrm{ET-}$\Delta + \textrm{CE-40km}$ & 2447 & 29924 & 395 & 2901 \\
        \textrm{ET-2L-mis + CE-40km} & 3743 & 36457 & 575 & 4301 \\
        ${\rm 1L}_{\rm EU}^{\rm 15\,km}$ + \textrm{CE-40km-mis} & 1557 & 19433 & 199 & 1603 \\
        ${\rm 1L}_{\rm EU}^{\rm 20\,km}$ + \textrm{CE-40km-mis} & 2077 & 23185 & 283 & 2116 \\
        ${\rm 1L}_{\rm EU}^{\rm 15\,km}{\rm \,HF}$ + \textrm{CE-40km-mis} & 903 & 13547 & 116 & 886 \\
        ${\rm 1L}_{\rm EU}^{\rm 20\,km}{\rm \,HF}$ + \textrm{CE-40km-mis} & 1392 & 17642 & 157 & 1229 \\
        \textrm{ET-}$\Delta$ & 35 & 914 & 2 & 28 \\
        ${\rm 1L}_{\rm EU}^{\rm 15\,km}$ + \textrm{CE-40km-mis + LIGO I} & 3171 & 28959 & 273 & 2144 \\
        \midrule
        \textrm{LVKI O5} & 38 & 599 & 1 & 1 \\
        \midrule
        \bottomrule
    \end{tabular}
    \caption{Number of detected BBH sources at the considered networks, with different cuts on the sky localization and relative error on the luminosity distance. The results are given for one year of data, with the duty cycle computed as in~\cite{Branchesi:2023mws}.}
    \label{tab:BBH_numbers_loc_wCE}
\end{table}

\begin{figure}[!tbp]
    \centering
    \includegraphics[width=.95\textwidth]{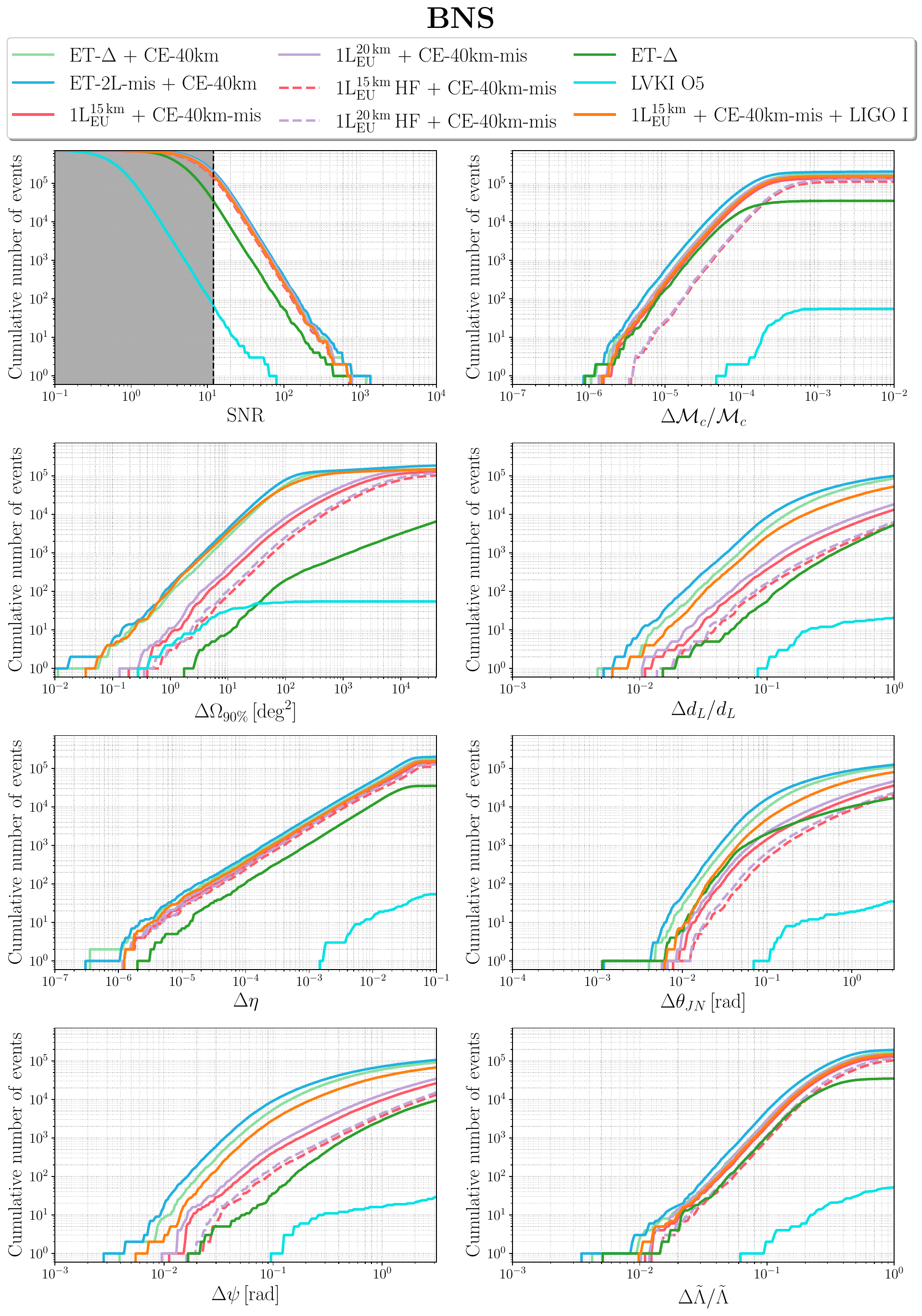}
    \caption{Cumulative distributions of the number of detections per year, for the SNRs and for the error on the parameters, for BNS signals,  for the various detector configurations considered.}
    \label{fig:PE_BNS_wCE}
\end{figure}

\begin{figure}[!tbp]
    \centering
    \begin{tabular}{c c c}
       \includegraphics[width=.31\textwidth]{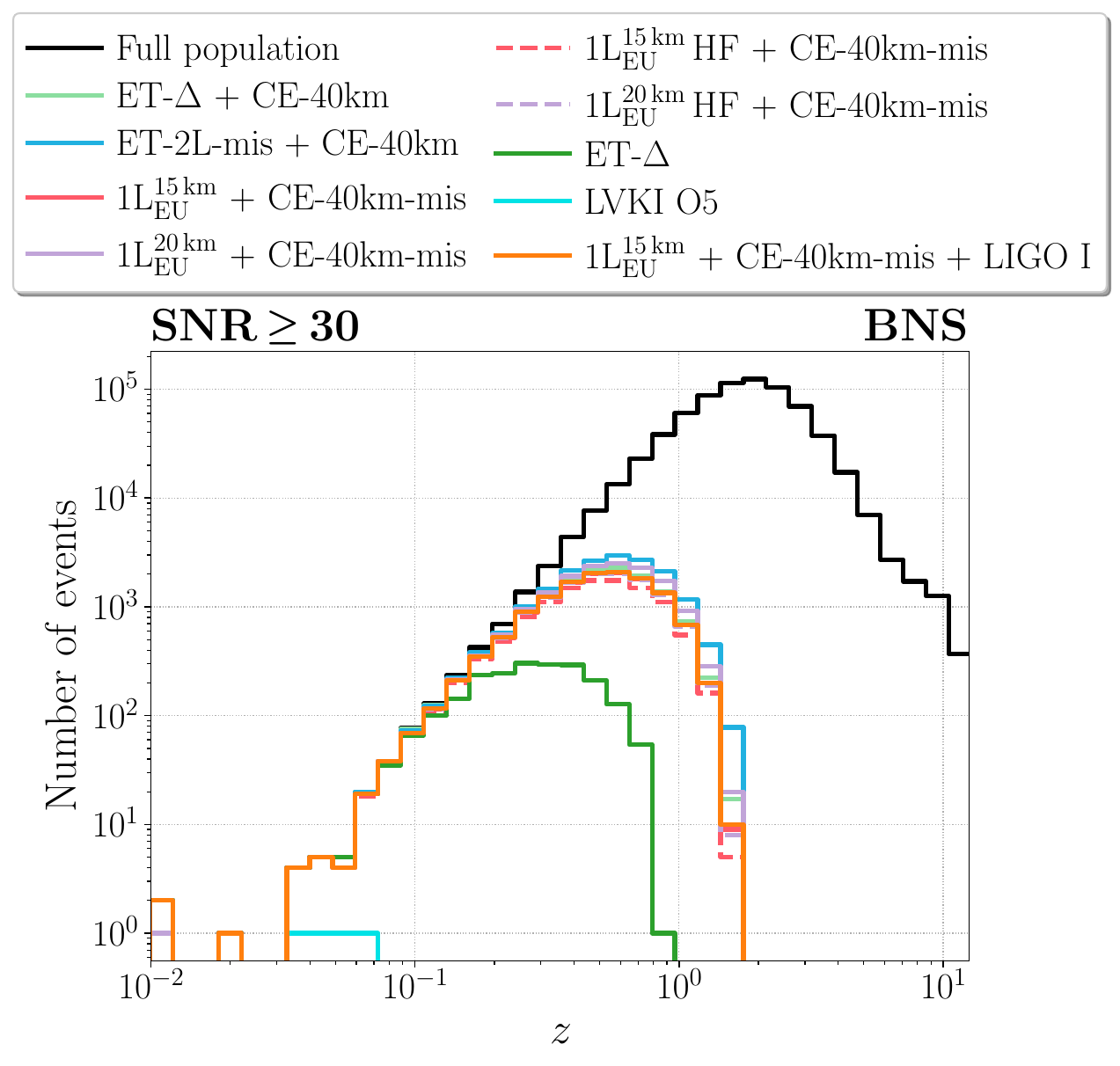}  & \includegraphics[width=.31\textwidth]{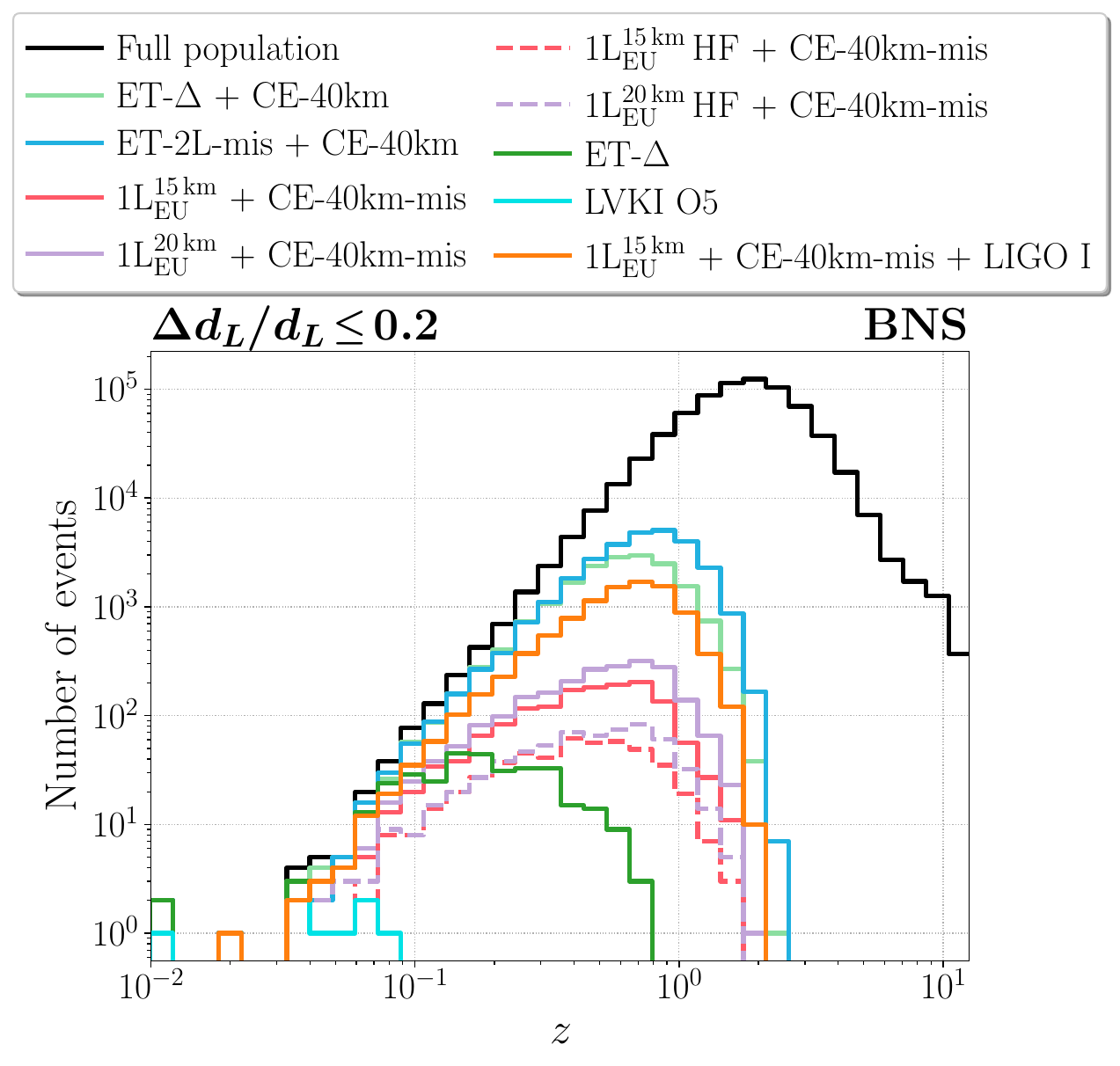} & \includegraphics[width=.31\textwidth]{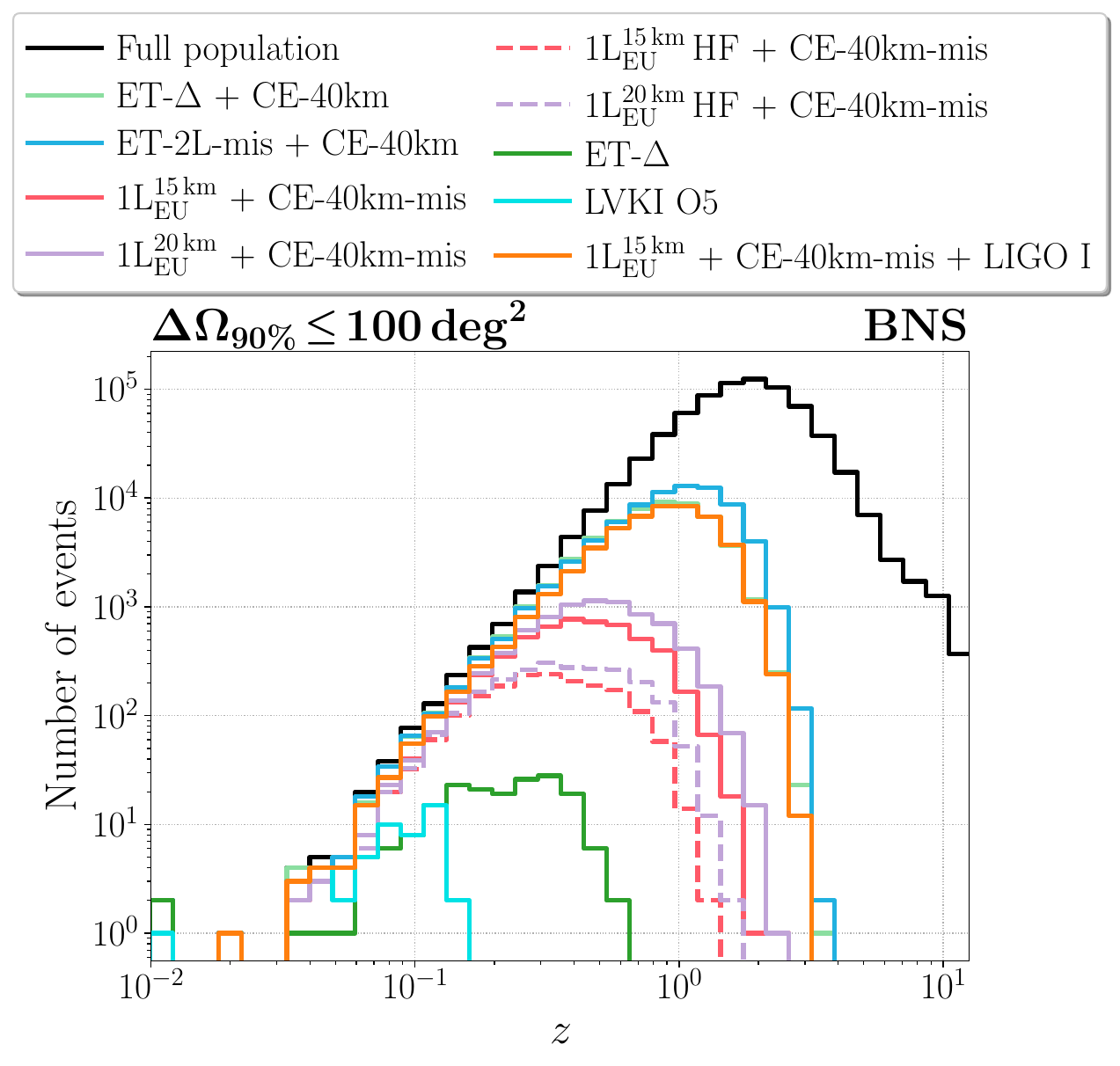} 
    \end{tabular}
    \caption{Redshift distribution of BNSs detected with ${\rm SNR} \geq 30$ (left column), relative error on the luminosity distance $\Delta d_L/d_L \leq 0.2$ (central column), or sky location $\Delta\Omega_{90\%} \leq 100~{\rm deg}^2$ (right column) for the various detector configurations considered.}
    \label{fig:histz_BNS_wCE}
\end{figure}

\begin{table}[!tbp]
    \centering
    \begin{tabular}{!{\vrule width 1pt}m{5.8cm}|S[table-format=4.0]|S[table-format=5.0]|S[table-format=4.0]|S[table-format=4.0]!{\vrule width 1pt}}
        \toprule
        \midrule
        \multicolumn{5}{!{\vrule width 1pt}c!{\vrule width 1pt}}{\textbf{BNS}}\\
        \midrule
        \hfil\multirow{3}{*}{\shortstack[c]{Detector\\configuration}}\hfill & \multicolumn{4}{c!{\vrule width 1pt}}{Detections with} \\
        & \multicolumn{2}{c|}{$\Delta\Omega_{90\%}\leq$} & \multicolumn{2}{c!{\vrule width 1pt}}{$\Delta d_L/d_L \leq$}\\
        \cmidrule(lr){2-5}
        & \multicolumn{1}{c|}{$10~{\rm deg}^2$} & \multicolumn{1}{c|}{$100~{\rm deg}^2$}  & \multicolumn{1}{c|}{$5\times10^{-2}$} & \multicolumn{1}{c!{\vrule width 1pt}}{$10^{-1}$}\\
        \midrule
        \textrm{ET-}$\Delta$ + \textrm{CE-40km} & 2427 & 54994 & 535 & 4100 \\
        \textrm{ET-2L-mis} + \textrm{CE-40km} & 3838 & 75828 & 1040 & 7949 \\
        ${\rm 1L}_{\rm EU}^{\rm 15\,km}$ + \textrm{CE-40km-mis} & 252 & 5388 & 59 & 347 \\
        ${\rm 1L}_{\rm EU}^{\rm 20\,km}$ + \textrm{CE-40km-mis} & 391 & 7852 & 90 & 506 \\
        ${\rm 1L}_{\rm EU}^{\rm 15\,km}{\rm \,HF}$ + \textrm{CE-40km-mis} & 72 & 1793 & 19 & 130 \\
        ${\rm 1L}_{\rm EU}^{\rm 20\,km}{\rm \,HF}$ + \textrm{CE-40km-mis} & 96 & 2403 & 25 & 154 \\
        \textrm{ET-}$\Delta$ & 8 & 184 & 8 & 52 \\
        ${\rm 1L}_{\rm EU}^{\rm 15\,km}$ + \textrm{CE-40km-mis + LIGO I} & 3161 & 49568 & 334 & 2499 \\
        \midrule
        \textrm{LVKI O5} & 31 & 51 & 0 & 1 \\
        \bottomrule
    \end{tabular}
    \caption{Number of detected BNS sources at the considered networks, with different cuts on the sky localization and relative error on the luminosity distance. The results are given for one year of data, with the duty cycle computed as in~\cite{Branchesi:2023mws}.}
    \label{tab:BNS_numbers_loc_wCE}
\end{table}

\clearpage

\subsection{Pre-merger alerts}\label{sect:premerger}

We next consider the ability of the various networks to provide an alert to electromagnetic telescopes before the merger.

\autoref{fig:premerger} shows how the SNR would accumulate in the various 3G  networks  considered, for an event with the characteristics of GW170817, as the  signal  sweeps up in frequency. For this event, which was very  close in distance, all world-wide 3G network configurations   considered have similar performances  (as long as  ET is in the LF+HF configuration) and rapidly reach  a very large SNR. A SNR larger than a detection threshold, say set at ${\rm SNR}=12$, is  reached in all case  about 10 hours prior to merger. Not surprisingly, loosing the LF instrument in ET has a significant effect on the pre-merger alert, since in this case the signal enters in the detector bandwidth only at higher frequencies, and therefore closer to merger.

\begin{figure}[tbp]
    \centering
    \includegraphics[height=.5\textwidth]{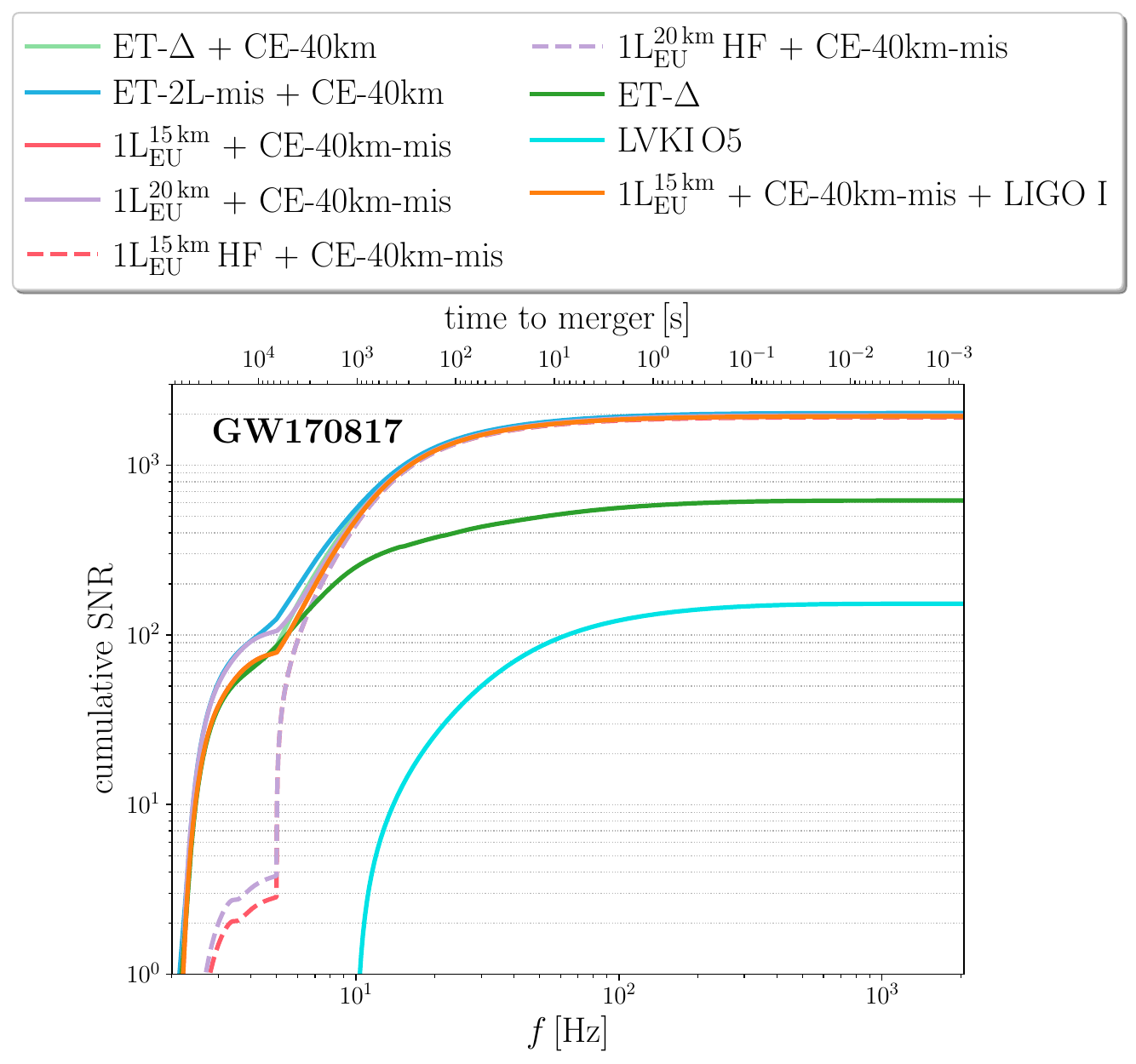}   
    \caption{Accumulation of the SNR in the various 3G network configurations, for a source with the properties of the BNS GW170817, as the signal sweeps up in frequency. The upper horizontal scale gives the corresponding time to merger. Observe that, for this system, the merger takes place at about 2~kHz, see e.g. Figure~2 of \cite{Iacovelli:2022bbs}.}
    \label{fig:premerger}
\end{figure}

\begin{table}[!tbp]
    \setlength\arrayrulewidth{.5pt}
    \renewcommand{\arraystretch}{1.07}
    \hspace{-1.6cm}
    \begin{tabular}{!{\vrule width 1pt}m{5.9cm}|c|c|S[table-format=3.0]|S[table-format=4.0]|S[table-format=5.0]|S[table-format=5.0]!{\vrule width 1pt}}
        \toprule
        \midrule
        \multicolumn{7}{!{\vrule width 1pt}c!{\vrule width 1pt}}{\textbf{BNS}}\\
        \midrule
        \hfil\multirow{2}{*}{\shortstack[c]{Detector\\configuration}}\hfill & \hfil\multirow{2}{*}{\shortstack[c]{Time before\\merger}}\hfill & \multirow{2}{*}{Orientation} & \multicolumn{4}{c!{\vrule width 1pt}}{Detections with $\Delta\Omega_{90\%}\leq$} \\
        \cmidrule(lr){4-7}
        & & & \multicolumn{1}{c|}{$10~{\rm deg}^2$} & \multicolumn{1}{c|}{$100~{\rm deg}^2$}  & \multicolumn{1}{c|}{$1000~{\rm deg}^2$} & \multicolumn{1}{c!{\vrule width 1pt}}{all sky}\\
        \midrule
        \multirow{6}{*}{\textrm{ET-}$\Delta + \textrm{CE-40km}$} & \multirow{2}{*}{30~min} & All $\Theta$ & 1 & 25 & 229 & 418 \\
        & & $\Theta\leq 15^\circ$ & 0 & 0 & 5 & 37 \\
        \cline{2-7}
        & \multirow{2}{*}{10~min} & All $\Theta$ & 11 & 234 & 1888 & 2493 \\
        & & $\Theta\leq 15^\circ$ & 0 & 9 & 64 & 233 \\
        \cline{2-7}
        & \multirow{2}{*}{1~min} & All $\Theta$ & 77 & 2140 & 22906 & 33042 \\
        & & $\Theta\leq 15^\circ$ & 3 & 76 & 790 & 3136 \\
        \midrule
        \multirow{6}{*}{\textrm{ET-2L-mis + CE-40km}} & \multirow{2}{*}{30~min} & All $\Theta$ & 0 & 41 & 307 & 875 \\
        & & $\Theta\leq 15^\circ$ & 0 & 0 & 9 & 82 \\
        \cline{2-7}
        & \multirow{2}{*}{10~min} & All $\Theta$ & 10 & 363 & 2521 & 4542 \\
        & & $\Theta\leq 15^\circ$ & 0 & 10 & 79 & 417 \\
        \cline{2-7}
        & \multirow{2}{*}{1~min} & All $\Theta$ & 101 & 2824 & 27880 & 42804 \\
        & & $\Theta\leq 15^\circ$ & 7 & 82 & 909 & 4057 \\
        \midrule
        \multirow{6}{*}{${\rm 1L}_{\rm EU}^{\rm 15\,km}$ + \textrm{CE-40km-mis}} & \multirow{2}{*}{30~min} & All $\Theta$ & 0 & 10 & 59 & 399 \\
        & & $\Theta\leq 15^\circ$ & 0 & 1 & 2 & 35 \\
        \cline{2-7}
        & \multirow{2}{*}{10~min} & All $\Theta$ & 0 & 41 & 373 & 2374 \\
        & & $\Theta\leq 15^\circ$ & 0 & 1 & 14 & 229 \\
        \cline{2-7}
        & \multirow{2}{*}{1~min} & All $\Theta$ & 8 & 229 & 2343 & 30344 \\
        & & $\Theta\leq 15^\circ$ & 0 & 9 & 98 & 2861 \\
        \midrule
        \multirow{6}{*}{${\rm 1L}_{\rm EU}^{\rm 20\,km}$ + \textrm{CE-40km-mis}} & \multirow{2}{*}{30~min} & All $\Theta$ & 0 & 12 & 96 & 757 \\
        & & $\Theta\leq 15^\circ$ & 0 & 0 & 0 & 69 \\
        \cline{2-7}
        & \multirow{2}{*}{10~min} & All $\Theta$ & 2 & 59 & 535 & 3637 \\
        & & $\Theta\leq 15^\circ$ & 0 & 2 & 25 & 342 \\
        \cline{2-7}
        & \multirow{2}{*}{1~min} & All $\Theta$ & 15 & 346 & 3268 & 35086 \\
        & & $\Theta\leq 15^\circ$ & 0 & 14 & 103 & 3318 \\
        \midrule
        \multirow{6}{*}{${\rm 1L}_{\rm EU}^{\rm 15\,km}{\rm \,HF}$ + \textrm{CE-40km-mis}} & \multirow{2}{*}{30~min} & All $\Theta$ & 0 & 0 & 0 & 42 \\
        & & $\Theta\leq 15^\circ$ & 0 & 0 & 0 & 9 \\
        \cline{2-7}
        & \multirow{2}{*}{10~min} & All $\Theta$ & 0 & 0 & 4 & 740 \\
        & & $\Theta\leq 15^\circ$ & 0 & 0 & 0 & 68 \\
        \cline{2-7}
        & \multirow{2}{*}{1~min} & All $\Theta$ & 1 & 9 & 111 & 21711 \\
        & & $\Theta\leq 15^\circ$ & 0 & 0 & 7 & 2017 \\
        \midrule
        \multirow{6}{*}{${\rm 1L}_{\rm EU}^{\rm 20\,km}{\rm \,HF}$ + \textrm{CE-40km-mis}} & \multirow{2}{*}{30~min} & All $\Theta$ & 0 & 0 & 0 & 42 \\
        & & $\Theta\leq 15^\circ$ & 0 & 0 & 0 & 9 \\
        \cline{2-7}
        & \multirow{2}{*}{10~min} & All $\Theta$ & 0 & 0 & 6 & 742 \\
        & & $\Theta\leq 15^\circ$ & 0 & 0 & 0 & 68 \\
        \cline{2-7}
        & \multirow{2}{*}{1~min} & All $\Theta$ & 2 & 16 & 174 & 21948 \\
        & & $\Theta\leq 15^\circ$ & 0 & 1 & 7 & 2037 \\
        \midrule
        \bottomrule
    \end{tabular}
    \caption{Number of BNS detected with $\rm SNR \geq 12$ and different cuts on the sky localization at 30~min, 10~min and 1~min prior to merger for the various detector configurations studied in this work. We further report the numbers both for all the sources in the catalog and the sub-sample of sources having an angle $\Theta \leq 15^\circ$, which can result more likely in coincident GRB detections.}
    \label{tab:premerger_time_angular_res_wCE}
\end{table}

\begin{table}[!tbp]
    \setlength\arrayrulewidth{.5pt}
    \renewcommand{\arraystretch}{1.07}
    \hspace{-1.6cm}
    \begin{tabular}{!{\vrule width 1pt}m{5.9cm}|c|c|S[table-format=3.0]|S[table-format=4.0]|S[table-format=5.0]|S[table-format=5.0]!{\vrule width 1pt}}
        \toprule
        \midrule
        \multicolumn{7}{!{\vrule width 1pt}c!{\vrule width 1pt}}{\textbf{BNS}}\\
        \midrule
        \hfil\multirow{2}{*}{\shortstack[c]{Detector\\configuration}}\hfill & \hfil\multirow{2}{*}{\shortstack[c]{Time before\\merger}}\hfill & \multirow{2}{*}{Orientation} & \multicolumn{4}{c!{\vrule width 1pt}}{Detections with $\Delta\Omega_{90\%}\leq$} \\
        \cmidrule(lr){4-7}
        & & & \multicolumn{1}{c|}{$10~{\rm deg}^2$} & \multicolumn{1}{c|}{$100~{\rm deg}^2$}  & \multicolumn{1}{c|}{$1000~{\rm deg}^2$} & \multicolumn{1}{c!{\vrule width 1pt}}{all sky}\\
        \midrule
        \multirow{6}{*}{\textrm{ET-}$\Delta$} & \multirow{2}{*}{30~min} & All $\Theta$ & 0 & 8 & 39 & 345 \\
        & & $\Theta\leq 15^\circ$ & 0 & 0 & 2 & 31 \\
        \cline{2-7}
        & \multirow{2}{*}{10~min} & All $\Theta$ & 0 & 28 & 125 & 1544 \\
        & & $\Theta\leq 15^\circ$ & 0 & 2 & 4 & 153 \\
        \cline{2-7}
        & \multirow{2}{*}{1~min} & All $\Theta$ & 4 & 79 & 372 & 7599 \\
        & & $\Theta\leq 15^\circ$ & 0 & 3 & 9 & 767 \\
        \midrule
        \multirow{6}{*}{${\rm 1L}_{\rm EU}^{\rm 15\,km}$ + \textrm{1CE-mis + LIGO I}} & \multirow{2}{*}{30~min} & All $\Theta$ & 0 & 10 & 59 & 399 \\
        & & $\Theta\leq 15^\circ$ & 0 & 1 & 2 & 35 \\
        \cline{2-7}
        & \multirow{2}{*}{10~min} & All $\Theta$ & 0 & 41 & 374 & 2374 \\
        & & $\Theta\leq 15^\circ$ & 0 & 2 & 14 & 229 \\
        \cline{2-7}
        & \multirow{2}{*}{1~min} & All $\Theta$ & 9 & 265 & 2848 & 30361 \\
        & & $\Theta\leq 15^\circ$ & 1 & 9 & 99 & 2861 \\
        \midrule
        \bottomrule
    \end{tabular}
    \caption{Continuation of \autoref{tab:premerger_time_angular_res_wCE}.}
    \label{tab:premerger_time_angular_res_wCE2}
\end{table}

For multi-messenger studies, however, we also need to get a good angular resolution before merger, in order to give  to  electromagnetic observatories at least some localization  information, and sufficiently early. 
\autoref{tab:premerger_time_angular_res_wCE} and \ref{tab:premerger_time_angular_res_wCE2} shows the
number of BNSs detected with $\rm SNR \geq 12$ and different cuts on the sky localization, at 30~min, 10~min and 1~min prior to merger, for the various network configurations considered.
We show separately the BNS detections where the  orbit has a generic inclination (as measured by the angle $\iota$ between the orbital angular momentum vector and the line-of-sight), and  those close to face-on, defined as those such that $\Theta \equiv {\rm min}\{\iota, 180^\circ - \iota\}$ is smaller  than $ 15^\circ$, which can result more likely in coincident $\gamma$-ray burst  (GRB) detections.

The results reflect the trends observed in \autoref{fig:PE_BNS_wCE}
for angular localization, with the (ET-2L-mis + CE-40km) and
(ET-$\Delta$ + CE-40km)  configurations significantly better than 
(1${\rm L}_{\rm EU}$ + CE-40km), and the latter  better than ET-$\Delta$ in isolation.  For instance, selecting the BNS that 10~min before merger are localized to better than $100\, {\rm deg}^2$ (and all inclination angles),
the best performance is from (ET-2L-mis + CE-40km), with 363 events/yr, followed by (ET-$\Delta$ + CE-40km)  with 234 events/yr, while (1${\rm L}_{\rm EU}^{15\, \rm km}$ + CE-40km) detects 41 events/yr, which remain
41 events/yr adding LIGO-India, i.e. for
(1${\rm L}_{\rm EU}^{15\, \rm km}$+ CE-40km + LIGO-I).
Note indeed that adding LIGO-India to the (1${\rm L}_{\rm EU}^{15\, \rm km}$ + CE-40km) network  
has little or  no effect on the premerger alerts at 30~min or 10~min before merger, and the addition of LIGO-India only raises appreciably the number of alerts much closer to merger; e.g. 
the rate of BNS localized to $100\, {\rm deg}^2$ (all inclination angles included) 1~min before merger raises from 229 to 265.

\subsection{Stochastic backgrounds}

Finally, we consider the sensitivity to stochastic GW backgrounds (SGWBs) of these configurations.

\begin{figure}[tbp]
    \centering
    \includegraphics[height=.6\textwidth]{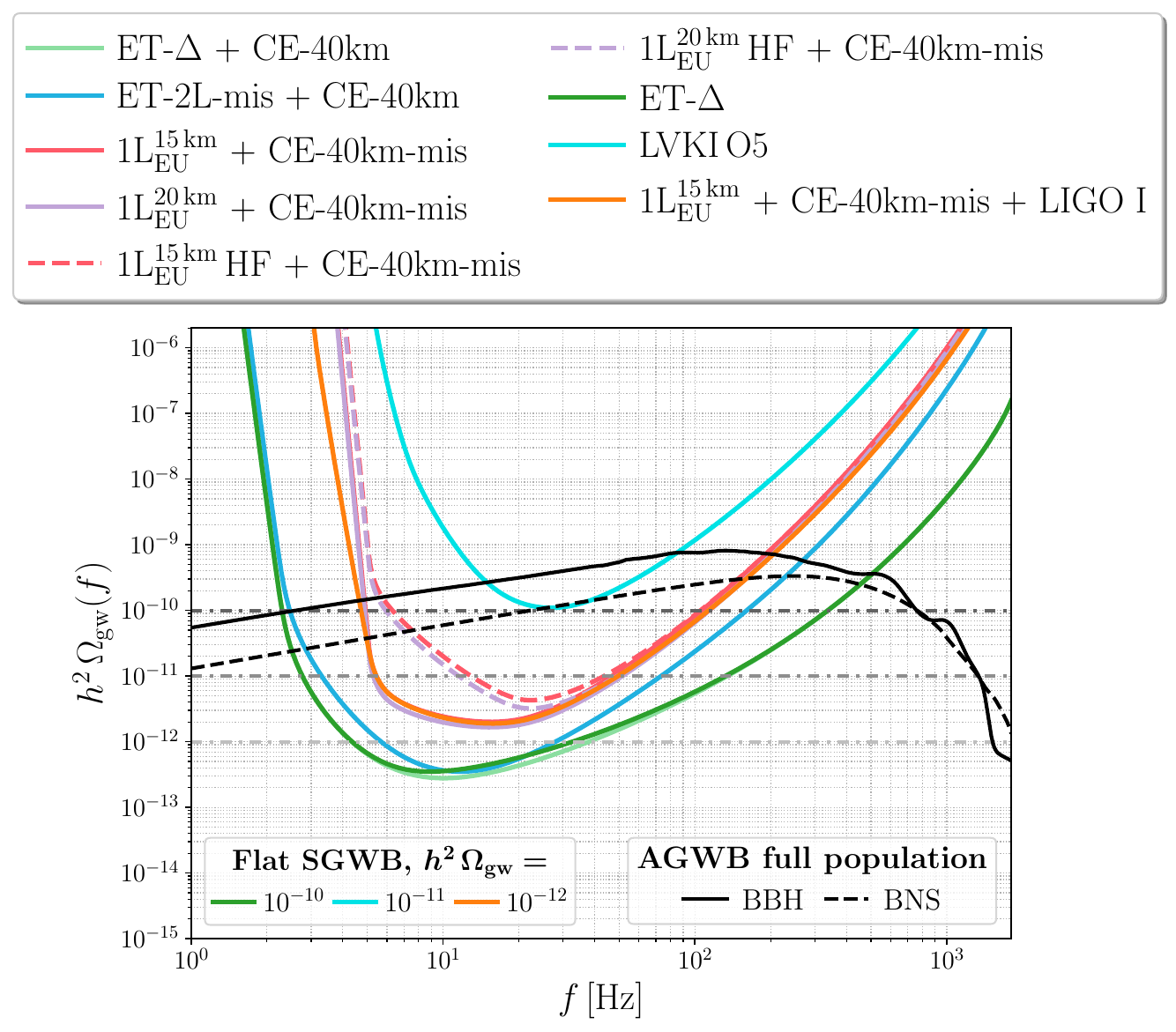} 
    \caption{Power-law integrated sensitivity curves for $h^2\,\Ogw(f)$  for the various configurations studied. These are computed assuming an observational time $T_{\rm obs}=1~{\rm yr}$ and an SNR threshold of 1. We further report some SGWB examples, namely a flat-in-frequency background with different amplitudes and the AGWB generated by our 1~yr catalogs of BBHs and BNSs. The PLS are computed  assuming uncorrelated noise among the detectors of a given network which, particularly for the triangle, might lead to a significant overestimate of the actual sensitivity to stochastic backgrounds, especially at low frequencies. Note that the curve for ET-$\Delta + \textrm{CE-40km}$ is mostly overlapped with that for ET-$\Delta $ alone; this reflects the fact that, because of the large intercontinental distance, the overlap reduction function suppresses the ET-CE correlations, so that, in ET-$\Delta + \textrm{CE-40km}$, most of the contribution comes from the correlation between the three colocated detectors composing the triangle. Similarly, (1${\rm L}_{\rm EU}^{15\, \rm km}$+ CE-40km)  and (1${\rm L}_{\rm EU}^{15\, \rm km}$+ CE-40km + LIGO-I) are indistinguishable.}
    \label{fig:PLSs}
\end{figure}

The sensitivity to stochastic background can  be expressed in terms of the power-law integrated sensitivity (PLS) introduced in \cite{Thrane:2013oya} (see also App.~A of ~\cite{Branchesi:2023mws} for conventions and definitions, and \cite{Belgacem:2025oom} for a recent discussion of its analytic properties).
We recall that the PLS is constructed considering power-law spectra of the form $\Omega_{\rm gw}(f)=\Omega_{\alpha} (f/f_*)^{\alpha}$, where, as usual, $\Ogw(f)=(1/\rho_c) {\rm d}\rho_{\rm gw}/{\rm d}\log f$, $\rho_{\rm gw}$ is the energy density in GWs, and  $\rho_c=3H_0^2/(8\pi G)$ is the critical density for closing the Universe~\cite{Maggiore:1999vm}; 
$f_*$ is a reference frequency, which can be chosen arbitrarily, since a change in $f_*$ can just be reabsorbed into a redefinition of $\Omega_{\alpha}$. To construct the PLS,  for each $\alpha$ one determines $\Omega_{\alpha}$ requiring that, performing the correlation between two detectors (or between more detectors, with the signal-to-noise ratio of the independent pairs summed in quadrature) with a given integration time, this spectrum be detectable at a given value of the signal-to noise ratio (${\rm SNR}$,  computed as in  \eq{Vol1_239} below); here we use the standard choice $({\rm SNR})_{\rm th}=1$ for the threshold in the signal-to noise ratio, and an integration time $T_{\rm obs}= 1$~yr. The PLS is then defined as the envelope of the family of spectra obtained in this way, as $\alpha$ is varied. By construction, 
every power-law spectrum that is tangent to the PLS at some frequency is detectable  with a detection threshold ${\rm SNR}_{\rm th}=1$,
and any power-law spectrum that, in some frequency range, is above the PLS, has ${\rm SNR} >1$.
\autoref{fig:PLSs} shows the  PLS of the various network configurations studied, expressed in terms of $h^2\,\Ogw(f)$
where $h$ is defined from 
$H_0=h \times (100~{\rm km\,s}^{-1}{\rm Mpc}^{-1})$.

\begin{table}[tbp]
    \centering
    \begin{tabular}{!{\vrule width 1pt}m{5.8cm}|S[table-format=3.1]|S[table-format=2.1]|S[table-format=1.1]|S[table-format=4.1]|S[table-format=3.1]!{\vrule width 1pt}}
        \toprule
        \midrule
        \hfil\multirow{3}{*}{\shortstack[c]{Detector\\configuration}}\hfill & \multicolumn{5}{c!{\vrule width 1pt}}{SNR for} \\
        & \multicolumn{3}{c|}{flat SGWB, $h^2\,\Omega_{\rm gw}=$} & \multicolumn{2}{c!{\vrule width 1pt}}{AGWB} \\
        \cmidrule(lr){2-4} \cmidrule(lr){5-6}
        & \multicolumn{1}{c|}{$10^{-10}$} & \multicolumn{1}{c|}{$10^{-11}$} & \multicolumn{1}{c|}{$10^{-12}$} & \multicolumn{1}{c|}{all BBH} & \multicolumn{1}{c!{\vrule width 1pt}}{all BNS}\\
        \midrule
        ET-$\Delta + {\rm CE-40km}$ & 358.8 & 35.9 & 3.6 & 828.7 & 234.9 \\
        ET-2L-mis + CE-40km & 285.3 & 28.5 & 2.9 & 698.7 & 199.7 \\
        ${\rm 1L}_{\rm EU}^{\rm 15\,km}$ + \textrm{CE-40km-mis} & 49.9 & 5.0 & 0.5 & 145.0 & 42.7 \\
        ${\rm 1L}_{\rm EU}^{\rm 20\,km}$ + \textrm{CE-40km-mis} & 60.1 & 6.0 & 0.6 & 175.2 & 51.6 \\
        ${\rm 1L}_{\rm EU}^{\rm 15\,km}$\,HF + \textrm{CE-40km-mis} & 23.3 & 2.3 & 0.2 & 80.4 & 24.1 \\
        ${\rm 1L}_{\rm EU}^{\rm 20\,km}$\,HF + \textrm{CE-40km-mis} & 31.2 & 3.1 & 0.3 & 107.2 & 32.1 \\
        ET-$\Delta$ & 285.2 & 28.5 & 2.9 & 621.6 & 174.5 \\
        ${\rm 1L}_{\rm EU}^{\rm 15\,km}$ + \textrm{CE-40km-mis + LIGO I} & 52.7 & 5.3 & 0.5 & 154.2 & 45.4 \\
        \midrule
        \bottomrule
    \end{tabular}
    \caption{SNRs at the various detector configurations for different stochastic background sources in an observational time $T_{\rm obs} = 1~{\rm yr}$. The columns 2, 3 and 4 show the results for a flat-in-frequency SGWB with different amplitudes; the fifth and sixth column for the AGWB from the superposition of  all the BBH (fifth column) and BNS (sixth column) signals present in our 1~yr catalogs. All results assume uncorrelated noise among the detectors of a given network.}
    \label{tab:SGWB_numbers_wCE}
\end{table}

An important caveat, when interpreting the PLS shown in 
\autoref{fig:PLSs}, is that these  are obtained assuming that the noise in the different detectors are uncorrelated. More precisely, the output in the $i$-th detector can be written as $s_i(t)=n_i(t)+h_i(t)$, where $n_i(t)$ is the noise of the $i$-th detector and $h_i(t)$ the projection of the GW signal onto the $i$-th detector. Let us recall that, if the noise between two different detectors is uncorrelated, the noise-noise correlator only grows with the observation time $T$ as $\sqrt{T}$ (as in a random walk) while the GW signal, which is correlated between the two detectors, grows as $T$ (see e.g. Section~7.8.3
of \cite{Maggiore:2007ulw} for more accurate discussion). In this way, correlating the output of two detectors and integrating for a long time, we can extract  a small signal from a much larger noise. As an order-of-magnitude estimate, correlating for a time $T$ the output of two detectors with a useful bandwidth $\Delta f$, we gain  a factor  of order $\sqrt{2 T\Delta f}$  in sensitivity (see eq.~7.246 of \cite{Maggiore:2007ulw}). Setting for instance $\Delta f$ of order  of a few hundred Hz, and $T=1$~yr, we then gain a factor ${\cal O}(10^5)$ in sensitivity. The curves in \autoref{fig:PLSs} have indeed been computed assuming uncorrelated noise, and integrating for 1~yr.
If, however, a noise is correlated, also the corresponding noise-noise correlator will grow as $T$ rather than $\sqrt{T}$. In this case, if in some frequency range a correlated noise is above a GW stochastic signal, we can integrate for as long as we want, but it will remain above the signal.  This problem is especially important for the triangle, with its three colocated interferometers. In particular,  correlated noise are expected  at low frequencies, because of Newtonian, seismic and magnetic noise, see 
\cite{Janssens:2021cta,Janssens:2024jln} and Section~5.4.1 of \cite{Branchesi:2023mws}. This means that, at low frequencies, the actual sensitivity of the triangle to stochastic GW backgrounds could actually be up to five orders of magnitude worse than that shown in \autoref{fig:PLSs}. The actual sensitivity is in fact difficult to compute, because it depends strongly on the level of correlated noise between the interferometers of the triangle configuration, which is difficult to predict before constructing and commissioning the detector. However, 
ref.~\cite{Janssens:2024jln}, reports correlations of several hundreds meters to a few kilometers
in underground seismic measurements from the ET candidate sites, and also elsewhere, in the frequency range from 0.01 to 40~Hz. From this, they project that the effect of correlated Newtonian noise from body waves on the sensitivity of ET in the triangle configurations will be severely degraded at least up to 20~Hz (while, well above 20~Hz, it eventually becomes negligible since it decreases rapidly with frequency).
Even if, to some extent, some correlated noise will  be present even between two well-separated L-shaped detectors, in this case the problem will be much less severe, and in particular the correlation of seismic noise will be negligible in the ET frequency band.

In \autoref{fig:PLSs} we also show, for comparison, the  current best estimate of the astrophysical GW background (AGWB)  obtained from the superposition of all BBH signals, and that obtained from  all BNS signals (computed as in Section~5.3 of \cite{Branchesi:2023mws}, see also \cite{Belgacem:2024ohp,Belgacem:2024ntv}). To guide the eye, we further draw 
horizontal lines corresponding to a  stochastic background where $h^2\,\Ogw$ is flat in frequency, and chosen to have the values  $10^{-10}$, $10^{-11}$, or $10^{-12}$.  Signals for which $h^2\,\Ogw$ is approximately constant in $f$, at least over the  range of frequencies explored by ground-based GW detectors, emerge rather naturally in cosmology, where typical spectra extend over many more decades in frequency, and the range explored by ground-based detectors is comparatively small; an explicit example is provided by some models for cosmic strings, see e.g. Figure~63 of \cite{Branchesi:2023mws}.
The  signal-to-noise ratio of a detector pair $(a,b)$ for a given stochastic background  signal is then given by
\be\label{Vol1_239}
{\rm SNR}= \[ 2T\int_0^{\infty}df\,  \(\frac{\Gamma_{ab}(f) S_h(f)}{S_{n,ab}(f)}\)^2 \]^{1/2}\, ,
\ee
where: $S_h(f)$ is the spectral density of the signal, related to the energy density  by $\Omega_{\rm gw}(f)=(4\pi^2/3H_0^2) f^3S_h(f)$; 
$S_{n,ab}=[ S_{n,a}(f) S_{n,b}(f) ]^{1/2}$, where $S_{n,a}(f)$ and $S_{n,b}(f)$ are the noise spectral densities of the two detectors labeled by $a,b$; and $\Gamma_{ab}(f)$ is their overlap reduction function,
\be\label{defGamma}
\Gamma_{ab}(f)=\int\frac{d^2\hatn}{4\pi}\sum_A F_a^A(\hatn)F_b^A(\hatn)\, e^{-i 2\pi f\hatn{\bf\cdot} (\vx_a-\vx_b)/c}\,, 
\ee
where $\vx_a$ and $\vx_b$ are the detector's location and $F_a^A(\hatn)$, $F_b^A(\hatn)$ the corresponding detector's pattern functions for the two polarizations $A=+,\times$, for a GW propagating in the direction $\hatn$; these are defined as $F_a^A(\hatn)=D^{ij}e_{ij}^A(\hatn)$ where $e_{ij}^A(\hatn)$ are the polarization tensors of the GW and $D^{ij}$ is the detector tensor, which depends only on the geometry of the detector; for an interferometers with arms in the $\hatu$ and $\hatv$ directions (not necessarily orthogonal),
$D_{ij}=(1/2)(u_iu_j-v_iv_j)$; see \cite{Maggiore:1999vm}, or
Sections~7.2 and 7.8.3 of \cite{Maggiore:2007ulw} for conventions and definitions. 

The results are shown in
\autoref{tab:SGWB_numbers_wCE} where, again, we have assumed uncorrelated noise, so the results for the triangle might actually be a gross overestimate of the actual sensitivity.

\section{Conclusions}

In this paper we have studied the performances of a global international  network made by a single 40~km CE detector in the US, together with a European third-generation GW detector. For the latter, we have examined
three different options: the two reference configurations for ET, namely a 10~km triangle and a 15-km 2L, and also a 1L configuration, in which in Europe there is only a single underground L-shaped detector with the ASD of ET. We have denoted these configuration, respectively,  as (ET-$\Delta$ + CE-40km), 
(ET-2L + CE-40km) and (1${\rm L}_{\rm EU}$ + CE-40km). In the latter case, we have considered both 15~km or 20~km arms for the 1${\rm L}_{\rm EU}$ detector.
We have further considered the effect of adding LIGO-India (at A$^\#$ sensitivity) to a network (1${\rm L}_{\rm EU}$ + CE-40km), as well as the effect of loosing the LF interferometer of ET in the (1${\rm L}_{\rm EU}$ + CE-40km) configuration. 

From the  point of view of the science case, the main conclusions that emerge from the comparison of these configurations  (restricting at first to ET in the HF+LF setting and neglecting for the moment the addition of LIGO-India to the
1${\rm L}_{\rm EU}$ + CE-40km network)
are as follows.

\begin{itemize}

\item For the detection horizons, the differences between these configurations are marginal, see \autoref{fig:All_horizons}. All these  networks will detect compact binaries up to similar distances, and over the same range of binary masses, from subsolar masses (a potential smoking-gun signature of primordial BHs) to BHs with masses of several thousands $\msun$, which provide the bridge toward supermassive BHs.

\item For  BBHs,   the results for parameter estimations are comparable among these networks, within factors of order 2,
with (ET-2L + CE-40km) always providing the best results, followed by (ET-$\Delta$ + CE-40km) and then 
by (1${\rm L}_{\rm EU}$ + CE-40km);
see  \hyperref[fig:PE_BBH_wCE]{Figures~\ref*{fig:PE_BBH_wCE}} and \ref{fig:histz_BBH_wCE}. This, combined with the results for detector horizons, implies that all aspects of the science case concerning BH astrophysics (demography, origin) are not significantly affected in the
(1${\rm L}_{\rm EU}$ + CE-40km) configuration, nor is the possibility of detecting and characterizing subsolar-mass primordial BHs or intermediate-mass BH. Similarly, the perspectives for extracting cosmological information by treating BHs as dark sirens (using either the correlation with galaxy catalogs or features in the mass function) are not significantly degraded, as well as the potential for studying deviations from General Relativity or the existence of compact objects from ringdown tests.

\item For BNSs  the situation is different. 
We see from \autoref{fig:PE_BNS_wCE} that the SNR distribution, and the reconstruction of masses and tidal deformabilities, are not very sensitive to the configuration. Therefore, all aspects of the science cased involving NS demography, as well as those related to the extraction of nuclear physics information from measurement of the tidal deformability, would be broadly preserved in the (1${\rm L}_{\rm EU}$ + CE-40km) configuration.

In contrast, the (1${\rm L}_{\rm EU}$ + CE-40km) configuration is one order of magnitude less performant than (ET-$\Delta$ + CE-40km) and 
(ET-2L + CE-40km) on the number of events with given angular localization and with given resolution on the luminosity distance. 
This could impact multi-messenger astronomy and the possibility of providing pre-merger alerts. For instance, we see from \autoref{tab:premerger_time_angular_res_wCE} and \ref{tab:premerger_time_angular_res_wCE2} that the number of BNSs detected 10 min before merger with a localization better than $100\, {\rm deg}^2$ (in our sample realization and in 1 yr, for our choice of duty cycle)  is
363 for (ET-2L + CE-40km), 234 for (ET-$\Delta$ + CE-40km), and only  41  for (1${\rm L}_{\rm EU}$ + CE-40km). From this point of view 
it should however be observed that, most  likely, electromagnetic telescopes will in any case  be able to perform a follow-up of  only a limited number of events per year, possibly providing the most significant bottleneck in this context. With  (1${\rm L}_{\rm EU}$ + CE-40km), however, we will also have a reduction in the number of the events with particularly good localization, as we see looking at the tails of the distribution in $\Delta\Omega_{90\%}$ in \autoref{fig:PE_BNS_wCE}. For instance, we can ask   what is the cut 
$\Delta\Omega_{\rm cut}$ that we can impose on angular resolution such that we remain with, say, 10 events with 
$\Delta\Omega_{90\%}<\Delta\Omega_{\rm cut}$ (we consider a set of 10 events because looking only at the very best event will be too sensitive to  statistical fluctuations associated to our sample realization); then, we find
that $\Delta\Omega_{\rm cut}\simeq 1\, {\rm deg}^2$ for 
(1${\rm L}_{\rm EU}$ + CE-40km), while 
$\Delta\Omega_{\rm cut}\simeq 0.2\, {\rm deg}^2$ for 
(ET-$\Delta$ + CE-40km) and 
$\Delta\Omega_{\rm cut}\simeq 0.1\, {\rm deg}^2$ for 
(ET-2L + CE-40km).

\item The sensitivity of the various configurations to stochastic backgrounds is shown in the left panel of \autoref{fig:PLSs}. The (1${\rm L}_{\rm EU}$ + CE-40km) configuration  improves significantly with respect to 2G detectors, but is clearly less performant than 
(ET-$\Delta$ + CE-40km) or 
(ET-2L + CE-40km). As discussed, however, the sensitivity of the configuration involving ET-$\Delta$ could be significantly degraded by correlated noise among the colocated detectors, especially at low frequencies.

\item Finally, we have found that, for coalescing binaries, a (1${\rm L}_{\rm EU}$ + CE-40km) international network provides results that are significantly better than  ET-$\Delta$ in isolation. On the one hand, this could have been expected because of the large inter-continental baseline of  (1${\rm L}_{\rm EU}$ + CE-40km); on the other hand,
this is non-trivial, considering  that ET-$\Delta$ is made of six interferometers, while  (1${\rm L}_{\rm EU}$ + CE-40km) only of three (two making up  1${\rm L}_{\rm EU}$, and one in CE, that only has a single interferometer per detector).

\end{itemize}

If we further add LIGO-India to the (1${\rm L}_{\rm EU}$ + CE-40km) network, the most significant improvement is in the angular localization (thanks to the long baseline with the other detectors), which in fact becomes comparable or even better to that of 
(ET-$\Delta$ + CE-40km); for BBH this can be seen from \autoref{tab:BBH_numbers_loc_wCE} that shows that, for the events localized within $1\, {\rm deg}^2$, (1${\rm L}_{\rm EU}$ + CE-40km) is in fact even better than 
(ET-$\Delta$ + CE-40km), with $3.2\times 10^3$  against 
$2.4\times 10^3$ events/yr. For BNS, \autoref{tab:BNS_numbers_loc_wCE} shows that, considering  events localized within $10\, {\rm deg}^2$, again $1\, {\rm deg}^2$, (1${\rm L}_{\rm EU}$ + CE-40km) is better than 
(ET-$\Delta$ + CE-40km), with again $3.2\times 10^3$  against 
$2.4\times 10^3$ events/yr.\footnote{The different cuts used for BNS and BBH reflect the fact that the BBH signals are in general stronger, because of the higher masses; the cuts on angular resolution are therefore chosen so to have comparable number of the events among the BBH and BNS cases.}

To sum up, these results indicate that, in an international context in which the US community builds  a single 40-km CE in the US, to be set in a network with ET, from the purely scientific point of view the European network should be based on the (ET-$\Delta$ + CE-40km) or
(ET-2L + CE-40km) configurations, that provide the best results. When correlated with a 40-km CE, the  ET-$\Delta$ and ET-2L options become quite comparable, with a general overall preference for ET-2L, as discussed in \cite{Branchesi:2023mws} (apart possibly for stochastic backgrounds, where however the effect of correlated noise could disfavor again the ET-$\Delta$ configuration).

However, 3G interferometers are enterprises of great technical complexity, requiring significant financial investment, manpower, etc. Should some  show-stopper  emerge, the result of our analysis indicate that, in the context of a network with a 40-km CE in the US (and only in the context of such a network), 
a European project based on just a single L-shaped
underground detector with the ASD of ET would be a viable (although temporary) back-up solution, that would
preserve a relevant part of the science case expected by the next generation of GW detectors. The addition of LIGO-India at  A$^\#$ sensitivity to the (1${\rm L}_{\rm EU}$ + CE-40km) network would
further significantly improve angular resolution, which would especially  benefit the aspects of the science case that are more sensitive to its, such as multi-messenger astronomy and standard-siren cosmology.
The price to pay in terms of scientific co-leadership, scientific independence and self-consistency is not evaluated in this paper. In particular, in order to guarantee a strategic and leading role for the ET project within the context of the global world-wide GW   research, if the ``L" geometry is chosen for ET, it is of paramount importance to implement ET in Europe through a network of detectors, each covering the whole ET frequency detection range, in order to minimise the impact of local disturbances, understand and control the noise sources, and benefit from the coherent analysis of the whole network at all frequencies.

\section*{Acknowledgements}

We thank Eugenio Coccia and Mario Martinez for very useful discussions that stimulated this study,  Harald L\"uck, Michele Punturo and B.~Sathyaprakash for useful comments on the manuscript, and Jerome Degallaix and Mikhail Korobko for their useful internal ET review.
The work of F.I., M.Mag. and N.M.   is supported by the
Swiss National Science Foundation (SNSF) grants 200020$\_$191957  and by the SwissMap National Center for Competence in Research.  E.B. and M.Mag. are supported by the SNSF
grant CRSII5$\_$213497. 
M.Mag. thanks for the hospitality the IFAE  in Barcelona, where part of this work was done.
The work of M.Manc. received support from the French government under the France 2030 investment plan, as part of the Initiative d'Excellence d'Aix-Marseille Universit\'e -- A*MIDEX AMX-22-CEI-02.
This work made use of the clusters Yggdrasil and Baobab at the University of Geneva.


\bibliographystyle{utphys}
\bibliography{myrefs.bib}

\end{document}